\documentclass[]{mn2e}

\usepackage{color}
\usepackage{amssymb}
\usepackage{amsmath}
\usepackage{graphicx}
\usepackage{appendix}
\usepackage{longtable}
\usepackage{url}

\title[High-mass end of $\phi(M_*)$]{The high mass end of the stellar mass function:  Dependence on stellar population models and agreement between fits to the light profile}
\author[Bernardi et al.]{\parbox{\textwidth}{M. Bernardi$^{1}$\thanks{E-mail: bernardm@sas.upenn.edu}, A. Meert$^{1}$, R. K. Sheth$^{1,2}$, J.-L. Fischer$^{1}$, M. Huertas-Company$^{3}$, C. Maraston$^{4}$, F. Shankar$^{5}$ \& V. Vikram$^{1}$} \vspace{0.4cm}\\
\parbox{\textwidth}{$^{1}$Department of Physics and Astronomy, University of Pennsylvania, Philadelphia, PA 19104, USA\\
$^{2}$The Abdus Salam International Center for Theoretical Physics, 
      Strada Costiera 11, 34151 Trieste, Italy\\
$^{3}$GEPI, Observatoire de Paris, CNRS, Univ. Paris Diderot;
Place Jules Janssen, 92190 Meudon, France\\
$^{4}$Institute of Cosmology and Gravitation, Dennis Sciama Building, Burnaby Road, Portsmouth PO1 3FX, UK\\
$^{5}$School of Physics and Astronomy, University of Southampton,
Southampton SO17 1BJ, UK\\}}

\begin{document}
 \date{Accepted .  Received ; in original form }

\maketitle

\label{firstpage}

\begin{abstract}
We quantify the systematic effects on the stellar mass function which arise from assumptions about the stellar population, as well as how one fits the light profiles of the most luminous galaxies at $z\sim 0.1$.  When comparing results from the literature, we are careful to separate out these effects. Our analysis shows that while systematics in the estimated comoving number density which arise from different treatments of the stellar population remain of order $\le 0.5$~dex, systematics in photometry are now about 0.1~dex, in contrast to some recent claims in the literature. Compared to these more recent analyses, previous work based on Sloan Digital Sky Survey (SDSS) pipeline photometry leads to underestimates of $\rho_*(\ge\! M_*)$ by factors of $3-10$ in the mass range $10^{11} - 10^{11.6}M_\odot$, but up to a factor of 100 at higher stellar masses. This impacts studies which match massive galaxies to dark matter halos.  Although systematics which arise from different treatments of the stellar population remain of order $\le 0.5$~dex, our finding that systematics in photometry now amount to only about 0.1~dex in the stellar mass density is a significant improvement with respect to a decade ago.  Our results highlight the importance of using the same stellar population and photometric models whenever low and high redshift samples are compared.  
\end{abstract}


\section{Introduction}
Our knowledge of galaxy formation and evolution has grown tremendously over the last two decades.  This knowledge comes from analysis of observed galaxy images, colors and spectra.  However, to quantify evolution, one must compare populations at different redshifts, so it is necessary to correct the observed colors or spectra to a common restframe waveband.  This is known as applying the $k+e$-correction.  In principle, this correction depends on the formation history of the stellar population which contributes to the observed light, so, in principle, it is different for each object.  Galaxy samples are now deep enough that $k+e$-corrections are necessary, and sample sizes are large enough that $k+e$-corrections can be a significant source of systematic error.  Since determining the appropriate $k+e$-correction for an object boils down to infering something about its star formation history, $k+e$-correction codes naturally also return an estimate of the stellar mass of the object.  Hence, the conversion from apparent brightness $\ell$ to luminosity $L$ in a fixed restframe waveband depends on the $k+e$-correction, so converting from $\ell$ to stellar mass $M_*$ does not require much more work.  This is one of the reasons why, over the last decade and a half, the emphasis has shifted to presenting results in terms of stellar mass $M_*$ rather than luminosity $L$.  Unfortunately, the conversion from $\ell$ to $L$ or $M_*$ is rather sensitive to how the stellar population was modelled -- what we will loosely refer to as the stellar mass-to-light ratio $M_*/L$ in what follows -- for which there is still no consensus (Bruzual \& Charlot 2003; {\tt Pegase}, Fioc \& Rocca-Volmerange 1997, 1999; Blanton \& Roweis 2007; Maraston et al. 2009; Conroy et al. 2009).  

During this same period, improvements in survey photometry have driven the development of more sophisticated algorithms for estimating the observed flux $\ell$.  A few years ago, Bernardi et al. (2010) noted that improving on SDSS pipeline photometry (sky subtraction issues, especially in crowded fields -- see e.g. Bernardi et al. 2007; Hyde \& Bernardi 2009; Blanton et al. 2011) indicated that the most luminous galaxies at $z\sim 0.1$ were more abundant than expected from the most commonly used parametrizations of the luminosity function.  Bernardi et al. (2013; hereafter B13) went on to show that fitting to single Sersic or two-component Sersic-Exponential profiles yielded substantially more light at the high mass, high luminosity end of the population.  However, here too, there is no consensus: e.g., D'Souza et al (2015) argue that their estimates of the total light in the brightest galaxies are substantially fainter than those of B13.  

Thus, for any given object, there are a wide variety of estimates of $\ell$ and $M_*/L$ available, with no agreement on which is correct.  As B13 have emphasized, this complicates the comparison of different authors' determinations of the stellar mass function $\phi(M_*)$.  Disagreement may be due to differences in how $\ell$ was estimated, or $M_*/L$, or both.  Similarly, agreement may be due to fortuitous cancellations of disagreements in both.  Since improvements in how $\ell$ is estimated can be made independently of how $M_*/L$ was determined, and vice versa, one may not argue that fortuitous cancellations indicate that $M_*$ is more robustly measured.  Therefore, it is useful and important to distinguish between these two possible sources of systematic agreement or disagreement.  

The main goal of the present paper is to quantify the current uncertainties on $\phi(M_*)$ with an emphasis on masses greater than $2\times 10^{11}M_\odot$.  This is the mass scale first identified by Bernardi et al. (2011) as being special:  Various scaling relations change slope at this scale, and this is thought to be related to a change in the assembly histories -- e.g. minor versus major dry mergers. It is also the mass scale where $\phi(M_*)$ starts to drop exponentially.  Therefore, large volumes are necessary to properly probe these objects.  E.g., a number of recent studies restrict attention to $z=0.06$ (e.g. the GAMA survey: Baldry et al. 2012; Kelvin et al. 2014; Taylor et al. 2015).  There are fewer than 100 such objects in the GAMA survey, so cosmic variance on these counts is an issue.  To reliably probe the high mass end, one must go to substantially larger survey footprints.  Even in the SDSS there are of order $10^3$ objects with $M_*\ge 2 \times 10^{11}M_{\odot}$ at $z\le 0.06$, so analyses which are restricted to small $z$, such as Weigel et al. (2016), do not provide reliable constraints on the abundance of massive galaxies.  

Section~\ref{photo} shows how $\phi(M_*)$ varies when $M_*/L$ is fixed and the method for determining $\ell$ is varied.  Section~\ref{ell} discusses the differences between SDSS pipeline photometry and more recent work.  Section~\ref{dsouza} argues that the $\ell$ estimates of D'Souza et al. (2015) are actually in good agreement with those of B13.  
In Section~\ref{sps}, the method for determining $\ell$ is fixed and only $M_*/L$ is varied.  This variation can result from different assumptions about the star formation histories in massive galaxies (Section~\ref{sp}) or the presence and effects of dust (Section~\ref{dust}).  Section~\ref{errs} discusses the impact of errors.  Section~\ref{recent} compares our findings with recent work.  
A final section summarizes our results.  

When necessary, we assume a spatially flat background cosmology with parameters $(\Omega_m,\Omega_\Lambda)=(0.3,0.7)$, and a Hubble constant at the present time of $H_0=70$~km~s$^{-1}$Mpc$^{-1}$.

\section{Systematic effects on $\phi(M_*)$ from photometry}\label{photo}
In what follows, we present results from the same set of $250,000$ objects which were analysed by Bernardi et al. (2010, 2013) with $14.5 < r_{\rm Petro} < 17.5$~mags in the $r-$band, selected from 4681~sq. degrees of the sky with a median redshift of about $z\sim 0.1$. We use as absolute magnitude of the Sun in the $r-$band $M_{r,\odot} = 4.67$.  We always estimate $\phi(M_*)$ using Schmidt's (1968) $V^{-1}_{\rm max}$ method, where $V_{\rm max}$ is always the value determined by B13.  We always bin in ${\rm log}_{10}(M_*/M_\odot)$, so we usually show ${\rm ln}(10)\,M_*\phi(M_*)$ in units of Mpc$^{-3}$dex$^{-1}$.  

We will compare several estimates of the apparent brightness of each galaxy.  The Sersic and SerExp fits used by B13 have since been published and made available online by Meert et al. (2015, 2016), for the full SDSS DR7 (Abazajian et al. 2009), so we often refer to them as the B13-Meert15 photometry.  Several alternative estimates of the apparent brightnesses for these objects are provided by Simard et al. (2011, hereafter Simard11).  Of the various estimates provided by the SDSS database, we will only use the DR9 $r-$band {\tt Model} and {\tt cModel} values (Ahn et al. 2012). The latter are known to be the least biased of the SDSS pipeline estimates (e.g. Bernardi et al. 2007; Bernardi et al. 2010).  

\subsection{Dependence on sky}\label{sky}
All of these estimates depend on correctly estimating (and removing) the contribution from the background sky. This is a major source of systematic bias in the flux determination especially for large nearby objects or those located in dense environments. The SDSS DR7 pipeline values for the sky are known to be systematically bright (Blanton et al. 2005; Bernardi et al. 2007; Hyde \& Bernardi 2009; Blanton et al. 2011; Meert et al. 2015; Fischer et al. 2017).  Therefore, the DR7 photometric parameters are based on oversubtracting the outer parts of large galaxies. This affects the photometry of those galaxies as well as of smaller and/or fainter objects in their vicinity.

Although the SDSS DR9 pipeline has improved sky subtraction around bright objects, these sky estimates are still significantly biased bright (Blanton et al. 2011). Blanton et al. argue that their sky estimates, based on fitting the masked sky background for each SDSS scan with a smooth continuous function, represent a substantial improvement over the standard SDSS catalog results and should form the basis of any analysis of nearby galaxies using the SDSS imaging data. Fischer et al. (2017) shows that this is not restricted to nearby galaxies: it is valid for all galaxies, and is particularly important for luminous galaxies located in dense environments. In addition, they show that the sky background estimates used in B13-Meert15 are in good agreement with those of Blanton et al. (2011) for all galaxies.  This simplifies the comparison with D'Souza et al. which follows, since they used DR9 objects with sky values of Blanton et al. (2011).  The agreement is also reassuring because PyMorph fits for the sky (which it treats as a constant across the image) at the same time that it fits the image of a galaxy.  Although one might worry that this produces degeneracies in the fitted values, running PyMorph with the sky fixed to that of Blanton et al. (2011) returns almost identical results, so this is not a concern.  For example, if {\tt PyMorph} is forced to fit a deVaucouleur model, the difference between fitting the sky as well, and using the sky of Blanton et al. (2011), has a scatter of just 0.02~mags around a median value of zero. (I.e., the slight 0.1~percent bias in the PyMorph sky estimate reported in Meert et al. 2013 is irrelevant.) A similar analysis of {\tt PyMorph}-Sersic and SerExp fits also shows no bias, but with a larger scatter ($\sim 0.05$~mags).  See Fischer et al. (2017) for more discussion.

\subsection{Dependence on apparent brightness $\ell$}\label{ell}
\begin{figure}
 \centering
 \includegraphics[width = 9cm]{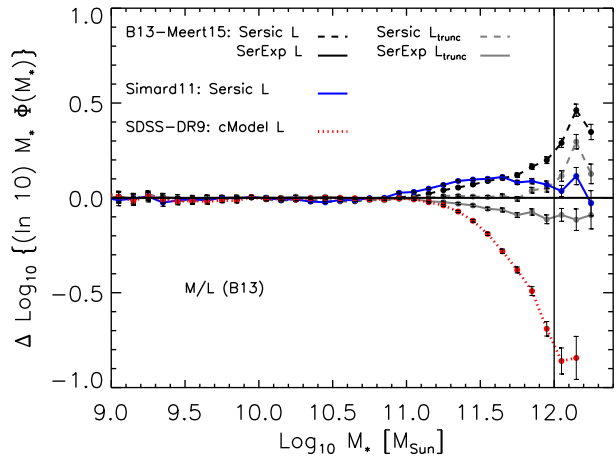}
 \caption{Comparison of six $\phi(M_*)$ estimates in which $M_*/L$ is fixed to the value used by B13, and the measured apparent brightness $\ell$, from which $L$ is derived, is varied.  To compare a large dynamic range, each $M_*\phi(M_*)$ is shown in units of $[M_*\phi(M_*)]_{\rm fid}$, where the fiducial value uses B13-Meert15 SerExp photometry.  All six estimates are in remarkable agreement below  $10^{11}M_\odot$; above this mass, the {\tt cModel} estimate becomes increasingly smaller than the others.}  
 \label{residLM}
\end{figure}

To set the stage, Figure~\ref{residLM} shows how $\phi(M_*)$ varies when $M_*/L$ is fixed (to the B13 values) and only $\ell$ is changed.  To better appreciate the trends over a large range in $M_*$, we do not show $\phi$ itself (see Figure~4 in B13), but $M_*\phi/(M_*\phi)_{\rm fid}$, where $(M_*\phi)_{\rm fid}$ uses the SerExp $\ell$ values of B13-Meert15. Three other choices of $\ell$ are also based on PyMorph:  one is the B13-Meert15 Sersic estimate, and the other two show the result of truncating the fitted (SerExp and Sersic) profiles at $7.5\times$ the half light radius when estimating the total magnitude (Fisher et al. 2017 shows that this closely approximates the SDSS algorithm).  The two other curves are based on the Simard11 Sersic and DR9 {\tt cModel} estimates.  (Replacing {\tt cModel} with {\tt Model} magnitudes gives almost identical results.)  The trends here are similar to those shown in Figure~1 of B13, except that there, the different $\ell$ estimates were compared, whereas here it is the impact on $\phi(M_*)$ which is shown.  

All six estimates are in remarkable agreement below  $10^{11}M_\odot$.  They become increasingly different above this mass.  B13 highlighted the fact that these systematic differences give rise to discrepancies in $\phi(M_*)$ determinations of more than 1~dex at $\ge\! 10^{12}M_\odot$.  However, they also noted that estimates based on fitting Sersic or SerExp profiles were in better agreement with one another than with {\tt cModel} magnitudes.   What is the real level of disagreement between these other estimates?  

We begin with a brief discussion of the fact that the B13-Meert values are based on integrating the fitted profile to infinity, whereas the SDSS truncates.  Briefly, (see Fischer et al. 2017 for a more detailed discussion), the effect of SDSS-like truncation is largest for single Sersic fits with large $n$ ($\sim 0.16$~mags for $n=8$).  There is a correlation between $L$ and $n$ -- large $L$ have larger $n$ -- so truncation matters more for large $L$.  However, the $n$-$L$ correlation is weak; at the largest $L$, the average $n$ is $6$ for which the correction is $\sim 0.12$~mags.  But because there is substantial scatter around the mean $n$, the net effect of truncation is about half than one would naively have expected.  Of course, truncation matters even less for the SerExp fits (see Figures 6 and 7 in Fischer et al. 2017).

Comparison of the black and grey curves in Figure~\ref{residLM} shows that truncation only matters above $\log_{10}(M_*/M_\odot) \ge 11.5$, where it introduces a systematic difference of 0.1~dex for SerExp, and slightly more for Sersic.  There is general agreement that the truncation scale should be of order $7-10\times$ the half light radius (D'Souza et al. 2015 use the lower value; Kelvin et al. 2012 use the higher value; the SDSS is in between).  Therefore, the actual uncertainty associated  with truncation is smaller than the 0.1~dex shown in Figure~\ref{residLM}.  This is much less than the difference with respect to {\tt SDSS-cModel}, which is our main point.  Henceforth, we ignore the  effects of truncation:  all PyMorph values we show are untruncated, since these are the values which are publically available. 

Discussion of the differences between Sersic or SerExp fits is complicated by the fact that B13-Meert15 and Simard11 are not in perfect agreement.  This is in part because the B13-Meert15 estimates are biased slightly bright (Meert et al. 2013; Bernardi et al 2014). This difference is mainly driven by the fact that the sky values assumed by Simard11 et al. are closer to the SDSS pipeline values (which are biased, see Section~\ref{sky}) than to PyMorph (see Fischer et al. 2017). The sky background difference leads to a median difference in the fitted Sersic magnitude of about $0.05-0.3$~mags (Meert et al. 2013; Fischer et al. 2017).

Potentially more important is the fact that recent work does not use the Simard11 photometric reductions `as is'.  Namely, Thanjavur et al. (2016) advocate using what they call `fiducial' values:  They choose the Simard11 Sersic value for most objects, and the Simard11 deVExp value (note that they do not advocate using the Simard11 SerExp fits at all) for the others.  (This choice depends on the quality of the fit to the surface brightness profile.)  Although we do not have their `fiducial' values, the solid curve in the top panel of their Figure~B1 lies slightly below the dashed curve, indicating that their `fiducial' values produce slightly smaller $M_*\phi(M_*)$ than one obtains if one simply uses the Simard11 Sersic values.  Note that the difference between these two only matters at large $M_*$, and is approximately the same as the amount by which the blue symbols in our Figure~\ref{residLM} differ from zero.  This is reassuring, since we believe that the B13-Meert15 SerExp $\ell$ values are our least biased estimates.  That is to say, results based on what each group believes to be its best estimate (B13-Meert15 SerExp and Thanjavur et al. `fiducial') are in even better agreement with one another than the 0.1~dex difference between the B13-Meert15 SerExp- and the Simard11 Sersic-based results shown in Figure~\ref{residLM}.  

Therefore, the most serious systematic difference is driven by the fact that {\tt cModel} magnitudes are much fainter than the others.  While B13 argue that this is a significant bias, recent work has questioned this conclusion.  E.g., D'Souza et al. (2015) do find a bias, but since their analysis of $\phi(M_*)$ resulted in fewer objects at high masses, they conclude that this bias was smaller than reported by B13.  
However, their conclusions are not based on a direct comparison of the two sets of photometry. As B13 emphasized, when using stellar masses to draw conclusions about photometry, it is important to ensure that differences in $M_*/L$ are not playing a role.  The next subsection shows that, once differences in $M_*/L$ have been accounted-for, the D'Souza et al. corrections to SDSS pipeline photometry are in remarkably good agreement with those of B13-Meert15.

\subsection{Comparison with D'Souza et al. (2015):  $\ell$}\label{dsouza}
Recently D'Souza et al. (2015) have used fits to stacked images to calibrate correction factors which they then apply to the SDSS {\tt Model} magnitudes of individual galaxies.  They perform these stacks separately for a number of different bins in stellar mass, where $M_*$ was obtained by combining the MPA-JHU $M_*/L$ values of the objects with the {\tt Model} magnitudes (because they could not stack on the corrected magnitudes, of course).  

\begin{figure}
 \centering
 \includegraphics[width=9cm]{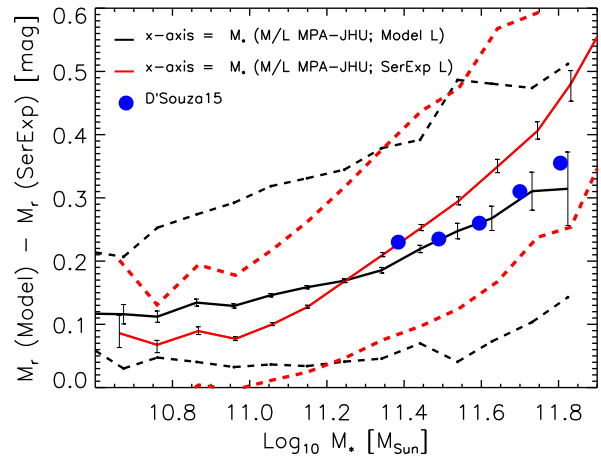}
 \caption{Corrections to SDSS pipeline photometry in the format of Figure~4 of D'Souza et al. (2015).  Filled symbols show the median correction applied by D'Souza et al. (2015) to the most massive objects, and black solid line shows that associated with the B13-Meert15 SerExp magnitudes for the same objects, both shown as a function of the $M_*$ estimate associated with MPA-JHU $M_*/L$ values (scaled to a Chabrier IMF) and SDSS {\tt Model} magnitudes.  Black dotted curves show the region which contains 68\% of the objects.  Since the D'Souza et al. corrections were computed as a function of MPA-JHU $M_*/L$ values and SDSS {\tt Model} magnitudes, this is the correct way to do the comparison, and the agreement is remarkable.
Red curves show {\tt Model}-SerExp as a function of the $M_*$ estimate associated with the same MPA-JHU $M_*/L$ values but SerExp magnitudes.  When shown this way, the implied corrections are larger, but comparing with the filled symbols, as D'Souza et al. did, is not justified.}
 \label{stacks2serexp}
\end{figure}

Figure~\ref{stacks2serexp} shows the median correction factors shown in their Figure~4, which are appropriate for the highest mass bins (redshift and concentration index) they consider -- the ones which are most relevant to our present study.  Notice that their values are brighter than the {\tt Model} magnitudes, in qualitative agreement with B13.  To see if the agreement is quantitative, the black solid curve shows the corresponding median values of the differences between the B13-Meert15 SerExp and {\tt Model} magnitudes for these same objects.  The agreement is remarkable.  

Why then did D'Souza et al. come to a different conclusion?  The reason is that their correction factor was calibrated in bins of $M_*$ in which {\tt Model} magnitudes were used to convert $M_*/L$ to $M_*$.  However, when they discussed the B13-Meert15 SerExp correction, they appear to be referring to some of the curves in Figure 1 of B13, which are binned in Ser magnitudes.  Since Ser is not the same as {\tt Model}, the objects in each $M_*$ bin are no longer the same, and so their discussion is not based on a direct comparison.
To illustrate this point, the red solid line shows the median {\tt Model}-SerExp correction as a function of SerExp- rather than {\tt Model}-based $M_*$.  Now the correction appears larger, but this is simply a consequence of the well-known fact that there is scatter between SerExp- and {\tt Model} magnitudes (Figure~1 in B13).
D'Souza et al. were incorrectly comparing the 0.3~mag correction associated with the blue symbols here with the 0.5~mag correction factor suggested by the red curve.  The correct comparison is with the black curve, and this shows that there is little or no difference.

The agreement is not quite as good as it seems, because D'Souza et al. obtained their corrections from truncated profiles: the inner component is truncated outside a radius equal to 7 half-light radius, while the outer component extends to infinity. In contrast, B13-Meert15 always integrate to infinity.  
If we were to account for this difference (see Figure 7 in Fischer et al. 2017), then the D'Souza et al. correction to {\tt Model} magnitudes would exceed that of B13-Meert15 (by about 0.05~mags; see also Figure~\ref{residLM} here and related discussion).  I.e., if anything, D'Souza et al. is brighter than B13-Meert15, not fainter.  This is perhaps not suprising; PyMorph SerExp forces the second component to be an exponential, whereas in D'Souza et al. allow it to be another Sersic, potentially including more light at large radii.

It is worth noting that, while the agreement between B13-Meert15 and Thanjavur et al. described in the previous section is perhaps not so surprising -- both are based on similar approaches to fitting the surface brightness profiles of individual galaxies, so they differ in the details of how the fit is actually done -- the agreement between the B13-Meert15 SerExp and the D'Souza et al. corrections shown in Figure~\ref{stacks2serexp} really is remarkable.  E.g., the stacking method must account for light lost to masked pixels, which are different for each member of the stack, and this will become increasingly important (and difficult!) at the high mass end.  This potential systematic is not an issue for approaches based on individual galaxies.  Another potential systematic is the sky, but we have already argued that because D'Souza et al. used the Blanton et al. (2011), rather than the SDSS pipeline, values for the sky, their sky values are in good agreement with those of B13-Meert15.

We now consider the scatter around the median.  The dashed lines show that this scatter (the region which encloses 68\% of the objects at each $M_*$) is slightly asymmetric, with an rms of approximately $\sim 0.2$~mags above $10^{11.3}M_\odot$.  It is similar to that shown in Figure~2 of D'Souza et al. (2015).  This rms is substantially larger than the quoted errors on either {\tt Model} or SerExp (Meert et al. 2013 quote errors of approximately 0.05~mags each; adding in quadrature yields 0.07~mags), so we conclude that most of it is intrinsic (because $0.07^2\ll 0.2^2$).  We return to this point in Section~\ref{DS}.  

\begin{figure}
 \centering
 \includegraphics[width=8.5cm]{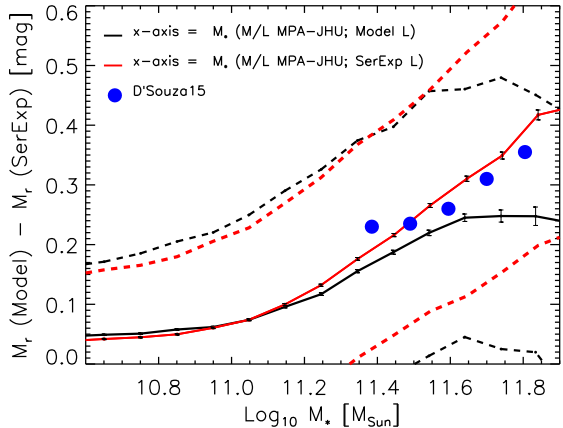}
 \caption{Same as previous figure but now for all galaxies in the mass bin.  }
 \label{stacks2serexpAll}
\end{figure}

Figure~\ref{stacks2serexpAll} shows a similar analysis of all galaxies -- not just those with high concentration indices.  Comparison with the previous panel shows that the median corrections here are smaller than they were for the high concentration galaxies.  (This is consistent with the fact that massive high concentration galaxies tend to be in crowded fields.)  Thus, for the galaxies with the biggest corrections, the median correction is the same as or smaller than made by D'Souza et al.

Based on the arguments above, we conclude that in fact there are no significant differences between the photometry of D'Souza et al. and B13.  Moreover, the agreement between the solid black line and the big blue symbols in Figure~\ref{stacks2serexp} indicate that {\tt cModel} is indeed biased by the amount claimed by B13.  Therefore, the fact that {\tt cModel} is an outlier in Figure~\ref{residLM} should not feature in discussions of the impact of photometry-related systematics on $\phi(M_*)$. The real level of systematic uncertainty is $\sim 0.1$~dex (see the other curves in Figure~\ref{residLM}).

\section{Dependence on stellar population modeling}\label{sps}
Having shown that systematics with more recent photometry contribute about 0.1~dex uncertainty at high masses (and that SDSS pipeline photometry is substantially fainter, in agreement with B13), we now study the systematics which arise from different assumptions about the stellar populations of the most massive galaxies.  Since we would like to separate the effects of stellar population modeling from those due to photometry, we must first account for the fact that not all the groups whose $M_*$ estimates we compare below used the same $\ell$ for estimating $L$.  To account for this, we multiply each group's $M_*$ value by $\ell_{\rm fiducial}/\ell$ where, for $\ell_{\rm fiducial}$, we use the $r$-band SerExp values provided by Meert et al. (2015).  Note that by using the ratio of the apparent brightnesses, we are making no change to the (stellar population dependent) $k+e$ values assigned by each group.  (We do account for the small differences which arise from the fact that not all groups used our fiducial cosmological parameters.)

\begin{figure*}
 \centering
 \includegraphics[width = 14cm]{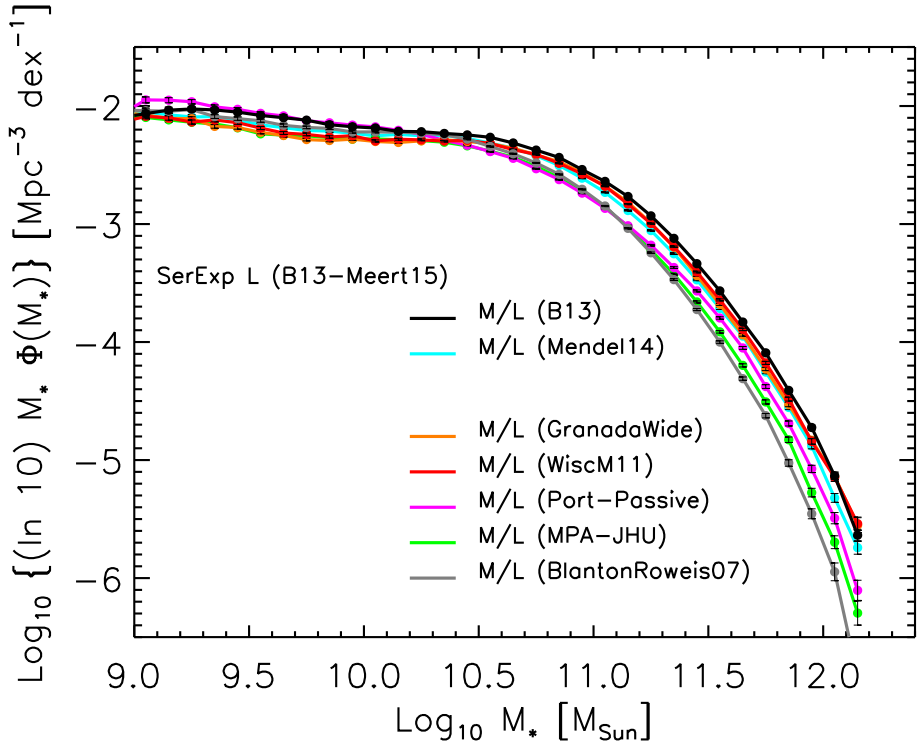}
 \caption{Comparison of a number of different determinations of $\phi(M_*)$ as labelled.  In all cases, the $M_*$ values were scaled to a Chabrier IMF and B13-Meert15 SerExp photometry, so the differences are entirely due to the stellar population modelling, fitting, and assumptions about dust in the galaxies. The Portsmouth-Passive estimate should be compared only at higher masses $M_* >\! 10^{11}M_\odot$ because a passive stellar population is not appropriate at smaller masses.}
 \label{compareML}
\end{figure*}

\begin{figure}
 \centering
 \includegraphics[width = 9cm]{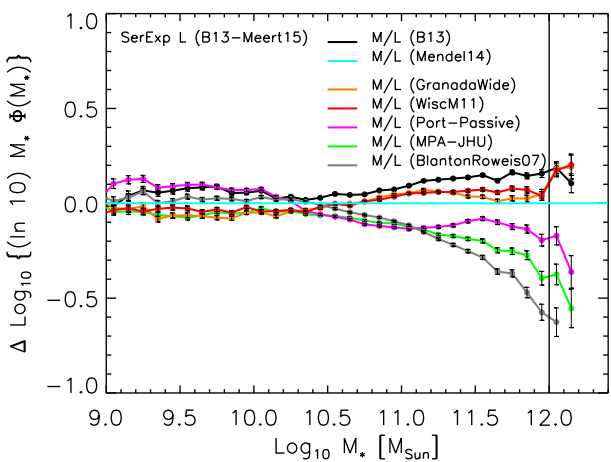}
 \caption{Same as previous figure, but now shown in units of $\phi_{\rm fid}$ for which we have used the cyan curve in the previous figure (i.e. SerExp photometry from B13-Meert15 and $M_*/L$ from dust-free Mendel14).  At $M_* >\! 10^{11}M_\odot$ the B13-based estimate is the largest, and the Blanton-Roweis is smallest.  The Wisconsin, GranadaWide and Mendel14 determinations are within about 0.1~dex of one another; they lie above the Portsmouth-Passive and MPA-JHU values. The Portsmouth-Passive estimate should be compared only at higher masses $M_* >\! 10^{11}M_\odot$ because a passive stellar population is not appropriate at smaller masses.}
 \label{residML}
\end{figure}

We have taken a number of stellar mass estimates from the SDSS database, {\tt www.sdss.org/dr12/spectro/galaxy/}, which all assumed a Kroupa (2001) initial mass function (IMF). These are:  
\begin{itemize}
 \item Spectro-photometric based estimates from the Portsmouth group (Maraston et al. 2013), which assume a passive stellar population (dust-free) that is expected to be appropriate for the most massive galaxies (Table {\tt stellarMassPassivePort}); 
 \item PCA-based estimates from the Wisconsin group (Chen et al. 2012) which are based on the stellar population synthesis models of Maraston \& Str\"omb\"ack (2011) and account for dust (Table {\tt stellarMassPCAWiscM11} -- for the high masses of most interest in our work, these are not very different from the models of Bruzual \& Charlot 2003);
 \item Estimates from the Granada group, which are based on the publicly available Flexible Stellar Population Synthesis code (FSPS, Conroy et al. 2009), and assume star formation occurs over a wide range of redshifts (Tables {\tt stellarMassFSPSGranWideDust} and {\tt stellarMassFSPSGranWideNoDust}); 
 \item Estimates from the MPA-JHU group (Table {\tt galSpecExtra}), which are based on the galSpec tool (building on the work of Kauffmann et al. 2003; Brinchmann et al. 2004; Tremonti et al. 2004).
\end{itemize}

There is no consensus on whether or not massive galaxies are dusty.  The Portsmouth-Passive estimates assume they are not.  This is based on Figure~8 of Thomas et al. (2013) who found very little dust in massive galaxies. On the other side, the MPA-JHU and Wisconsin estimates assume they are.  The GranadaWide group allow for both possibilities.

We have also included older estimates from Blanton \& Roweis (2007) and a particularly simple estimate used by Bernardi et al. (2010, 2013), in which $M_*/L$ is a linear function of color (see Eq. 6 in Bernardi et al. 2010, which is motivated by Bell et al. 2003 who used the {\tt Pegase} models), primarily for comparison with previous work.  Finally, we include more recent estimates from Mendel et al. (2014; their Tables~3 and~5); they provide dusty and dust-free $M_*/L$ estimates, and do not allow for multiple bursts.

When computing stellar masses there is also no consensus on whether or not one should allow for star formation histories with multiple bursts.  The MPA-JHU estimates assume bursty histories, whereas the Portsmouth-Passive and Mendel et al. models do not.  Mendel et al. exclude bursts because Gallazzi \& Bell (2009) showed that including bursty SPS models can lead to a systematic underestimate of M/L by up to 0.1 dex for galaxies whose star-formation histories are generally smooth. This is likely to be the case for the most massive galaxies.

As we show in Section~\ref{Moustakas}, dust and bursty-ness contribute significantly to the error budget at the high mass end of the stellar mass function (e.g. Moustakas et al. 2013; Mendel et al. 2014).

There is also no consensus on the shape of the IMF.  We have (arbitrarily) chosen to correct all $M_*$ values to a Chabrier IMF (Chabrier 2003).  No correction is necessary for the Blanton-Roweis, B13, and Mendel14 estimates.  However, the Portsmouth, Wisconsin, Granada and MPA-JHU estimates assumed a Kroupa IMF (Kroupa 2001); to convert to Chabrier we simply decrease the quoted (Kroupa) $M_*$ values by 0.05~dex (see, e.g., Table~2 in Bernardi et al. 2010).  

Appendix~A compares the $M_*/L$ estimates from these different analyses.  It shows that, above $10^{11}M_\odot$, these differences are of order 0.05~dex or larger, provided we compare dusty or dust-free models separately (but not dusty with dust-free).  However, there is no consensus on the effects of dust.  For Mendel14 allowing for dust decreases the inferred $M_*/L$ by 0.05~dex or more (Figure~\ref{mendelDust}), but for WiscM11 allowing for dust increases it by about 0.06~dex (Figure~13 in Chen et al. 2012).  The impact of dust on the GranadaWide $M_*/L$ values is more complicated (see Appendix~A for details).  The next subsection shows how these differences impact $\phi(M_*)$.

\subsection{Effect of stellar population model}\label{sp}
Figure~\ref{compareML} compares a number of estimates of $\phi(M_*)$ which are based on the publically available $M_*$ values listed earlier.  Below $10^{10.5}M_\odot$ most estimates of $\phi(M_*)$ differ by of order 0.1~dex (we argue later that this should not be used to argue that $M_*/L$ in low mass galaxies is well understood!).  The Portsmouth-Passive estimate should be compared only at higher masses $M_* >\! 10^{11}M_\odot$ because a passive stellar population is not appropriate at smaller masses.  Using their star-formation model (Table {\tt stellarMassStarformingPort}) lowers the value of $\phi(M_*)$ (we have not shown this in the figure because it is not a good choice for the high masses of most interest in our work).

At larger masses, the differences can be much larger.  To show this more clearly, Figure~\ref{residML} presents these curves normalized by $\phi_{\rm fid}$, for which we use the one associated with the B13-Meert15 SerExp photometry and the $M_*/L$ value of Mendel14.  The results are bracketed by the curves associated with the Blanton-Roweis (lowest) and B13 (uppermost) $M_*/L$ values.  Whereas the Blanton-Roweis values are known to be inappropriate for the most massive galaxies (see discussion in Bernardi et al. 2010), the B13 prescription is slightly ad hoc (e.g., the $k+e$ corrections and $M_*/L$ values are not derived self-consistently).  Nevertheless, even for the better motivated models, these are not small offsets ($\ge\! 0.5$~dex or more at $10^{12}M_\odot$), especially in view of the fact that these massive galaxies are expected to have the simplest (single burst) stellar populations.

\subsection{Effect of dust}\label{dust}
\begin{figure}
 \centering
 \includegraphics[width = 9cm]{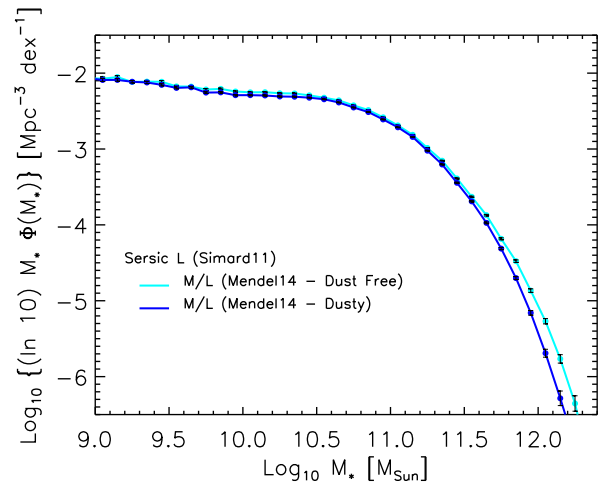}
 \caption{Comparison of $\phi(M_*)$ associated with the dust-free and dusty $M_*/L$ values of Mendel et al. (2014). Only in this Figure $M_*$ is computed using Simard11 Sersic photometry as this eases comparison with Figure B1 in Thanjavur et al. (2016).}
 \label{compareMendel}
\end{figure}

Some of these differences arise from making different assumptions about dust in these galaxies.  Figure~\ref{compareMendel} compares $\phi(M_*)$ associated with the dusty and dust-free $M_*/L$ values of Mendel et al. (2014); the difference  is entirely a consequence of the differences shown in Figure~\ref{mendelDust}.  The dust-free models have larger $M_*/L$, so they result in larger $\phi(M_*)$; at $10^{12}M_\odot$ the difference is about 0.4~dex. (For ease of comparison with Figure B1 in Thanjavur et al. (2016), in this Figure $M_*$ is computed using Simard11 Sersic photometry.)

\begin{figure}
 \centering
 \includegraphics[width = 9cm]{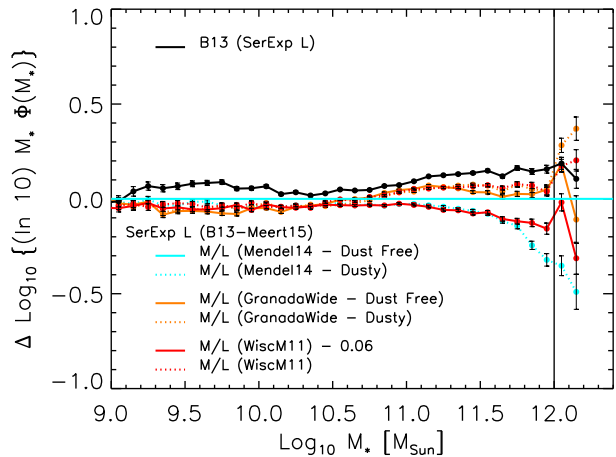}
 \caption{Same as Figure~\ref{residML}, but now comparing $\phi(M_*)$ estimates in which galaxies are assumed to be dusty (dotted) or dust-free (solid).  Since only dusty WiscM11 $M_*$ values are available, the solid curve for this case is based on making the $M_*/L$ values 0.06~dex smaller (see text for details).}  
 \label{residDust}
\end{figure}

To see if this level of discrepancy is the same in other models, we have compared dusty and dust-free estimates of $\phi(M_*)$ in the GranadaWide and Wisconsin models as well.  Figure~\ref{residDust} shows the results:  solid curves represent analyses which assume the galaxies are dust-free; dotted curves assume the galaxies are dusty.  (In fact, only dusty WiscM11 $M_*$ values are available.  However, Figure~13 of Chen et al. 2012 shows that assuming the galaxies are dust-free decreases the WiscM11 $M_*/L$ values by about 0.06~dex.  So, the WiscM11-NoDust $\phi(M_*)$ curves in Figure~\ref{residDust} were approximated by simply shifting their $M_*$ values downwards by this amount.)

We noted before that there is not even consensus on the sign of the effect, and this is reflected in the $\phi(M_*)$ estimates.  In the Wisconsin models, allowing for dust increases the inferred $M_*/L$ and hence $\phi(M_*)$; for Mendel14, accounting for dust reduces $\phi(M_*)$ by a similar amount.  As a result, the dust-free Mendel14 estimate happens to be reasonably like the WiscM11 dusty estimate, and vice-versa.  On the other hand, dust has little effect on $\phi(M_*)$ of the GranadaWide models (the effect on $M_*/L$ is actually more complicated; see Appendix~A).

Before moving on, it is worth noting that the difference between the B13 $\phi(M_*)$ and the dust-free Mendel14 curve is smaller than the $\sim 0.3$~dex reported in Thanjavur et al. (2016).  This is because their analysis was based on the dusty Mendel14 models -- the dotted cyan curve in Figure~\ref{residDust}.  This shows that the differences they report are largely due to differences in $M_*/L$ which are largely due to different assumptions about dust in these galaxies.  

\begin{figure}
 \centering
 \includegraphics[width = 9cm]{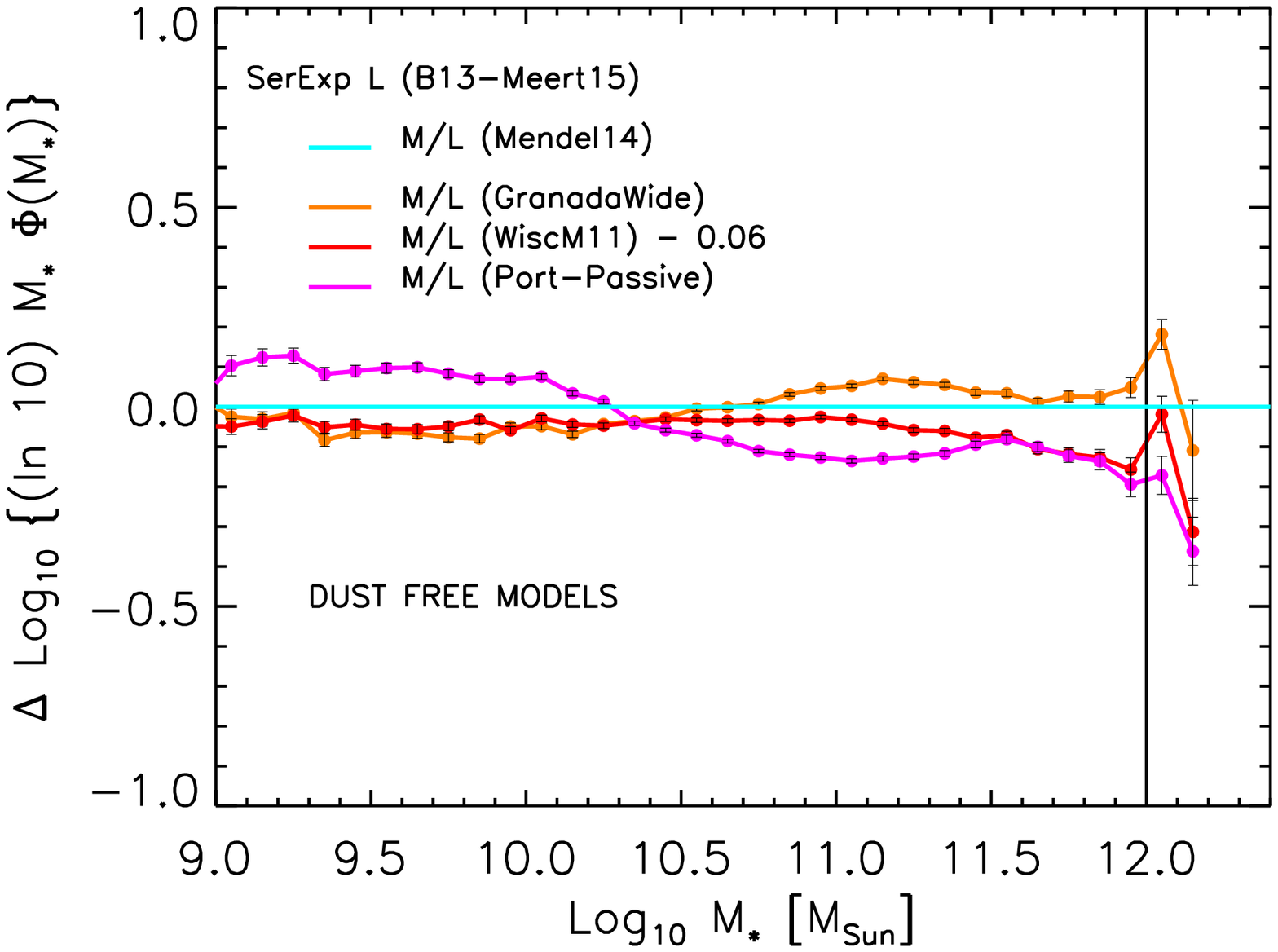}
 \includegraphics[width = 9cm]{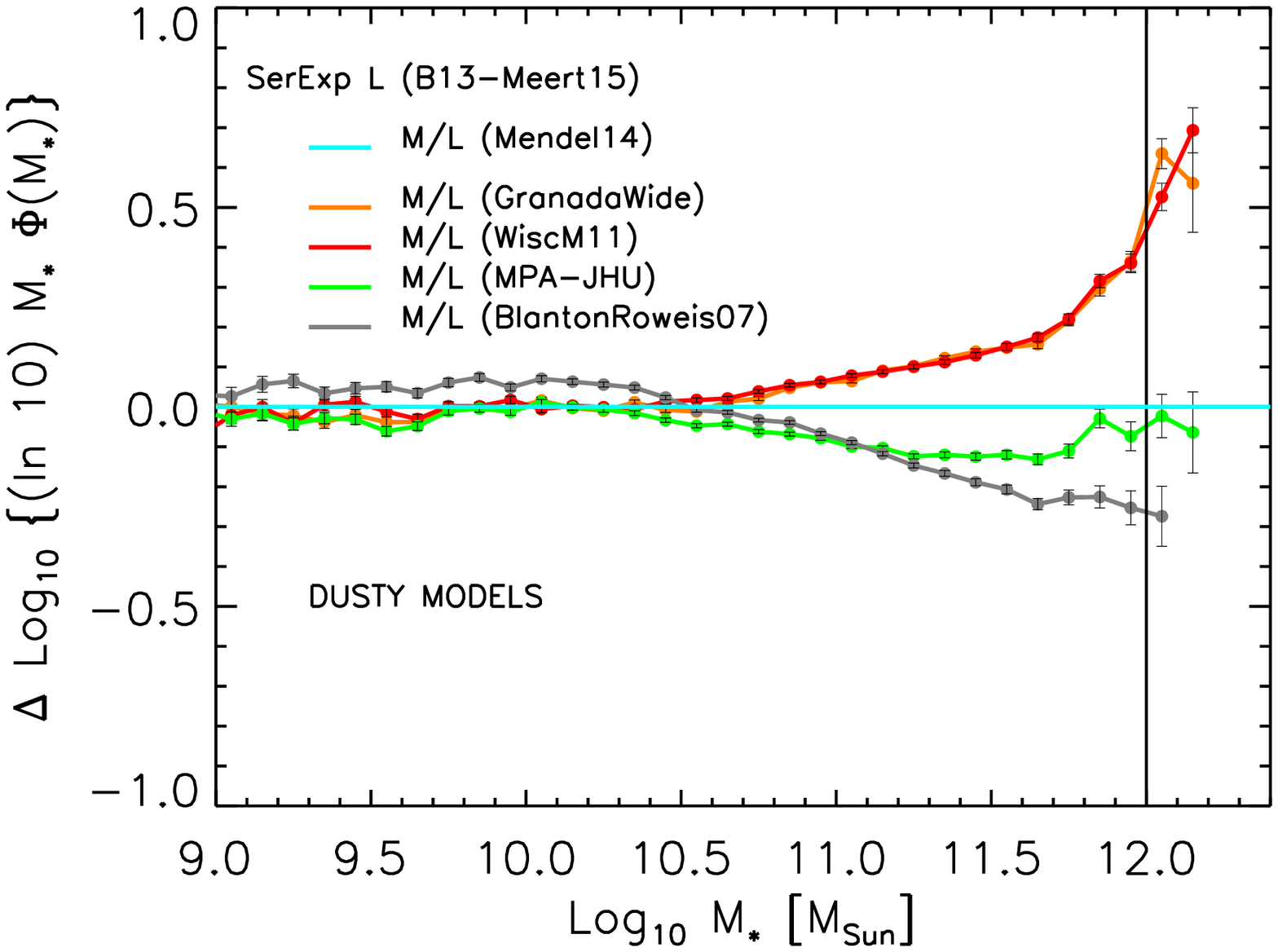}
 \caption{Same as previous figure, but now comparing dust-free models (top) and dusty models (bottom) separately.  }
 \label{dustNoDust}
\end{figure}

Until we know whether dusty or dust-free models are correct, we thought it prudent to show the level of systematics on $\phi(M_*)$ if we compare the two sets of models separately.  This is done in Figure~\ref{dustNoDust}.  The systematic differences on $\phi(M_*)$ between dust-free models are slightly smaller than for dusty models:  $\le 0.3$~dex compared to $\sim 0.4$~dex at $10^{12}M_\odot$.

\begin{figure}
 \centering
 \includegraphics[width = 9cm]{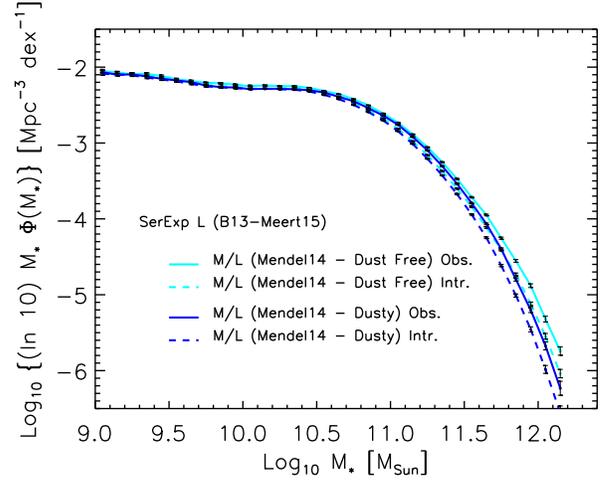}
 \caption{Comparison of the measured (solid) and intrinsic (dashed) stellar mass functions for the dusty and dust-free models of Mendel14 (with SerExp photometry of B13-Meert15).  Measurement errors matter little below $10^{11}M_\odot$, but enhance $\phi(M_*)$ by more than 0.3~dex at $M_* \sim 10^{12}M_\odot$.}
 \label{errsSDSS}
\end{figure}

\subsection{Effect of measurement errors}\label{errs}
So far we have studied how the measured $\phi(M_*)$ depends on $M_*/L$ and $\ell$.  However, the estimated $M_*$ values are noisy, with statistical (not systematic!) uncertainties of $\sim 0.1$~dex.  While these matter little at the lower masses where $\phi(M_*)$ is relatively flat, they modify $\phi(M_*)$ substantially at the high mass end of most interest here.  
If the intrinsic $M_*\phi_{\rm int}(M_*)$ is approximately flat at low masses and falls as $\exp[-(M/M_*)^\beta]$ at high masses, then errors which are Gaussian in $\log_{10}(M_*)$ make 
\begin{eqnarray}
 \frac{\phi_{\rm obs}}{\phi_{\rm int}} - 1 &\approx& 
 \frac{(\beta\, \ln(10)\,\sigma_{\log_{10} M})^2}{2} \, \left(\frac{M_*}{M_{*{\rm -char}}}\right)^\beta\,\nonumber\\
 &&\qquad \times \quad \left[\left(\frac{M_*}{M_{*{\rm -char}}}\right)^\beta - 1\right].
 \label{b10eq10}
\end{eqnarray}
(equations~10 and~11 of Bernardi et al. 2010).

Notice that errors matter little below a characteristic mass $M_{*{\rm -char}}$; this is not surprising, since the flatness of $M_*\phi_{\rm int}(M_*)$ below $M_{*{\rm -char}}$ means that errors move objects between mass bins, but because all bins have the same height, there is no net change to $M_*\phi_{\rm int}(M_*)$.  Fits which constrain $\beta=1$ tend to return $M_*\sim 10^{11}M_\odot$ (e.g. D'Souza et al. 2015; Thanjavur et al. 2016), so errors only matter at higher masses.  (E.g., at $10^{12}M_\odot$, errors of order $\sigma_{\log_{10} M}=0.1$~dex make $\log_{10}(\phi_{\rm obs}/\phi_{\rm int}-1)\approx 0.4$~dex.)  This means that the 0.1~dex agreement between various determinations of $\phi(M_*)$ at lower masses should not be used to argue that $M_*/L$ in low mass galaxies is well understood (e.g. Thanjavur et al. 2016). As Appendix~A shows, this agreement is hiding large systematic differences between different groups.  

\begin{figure}
 \centering
 \includegraphics[width = 9.cm]{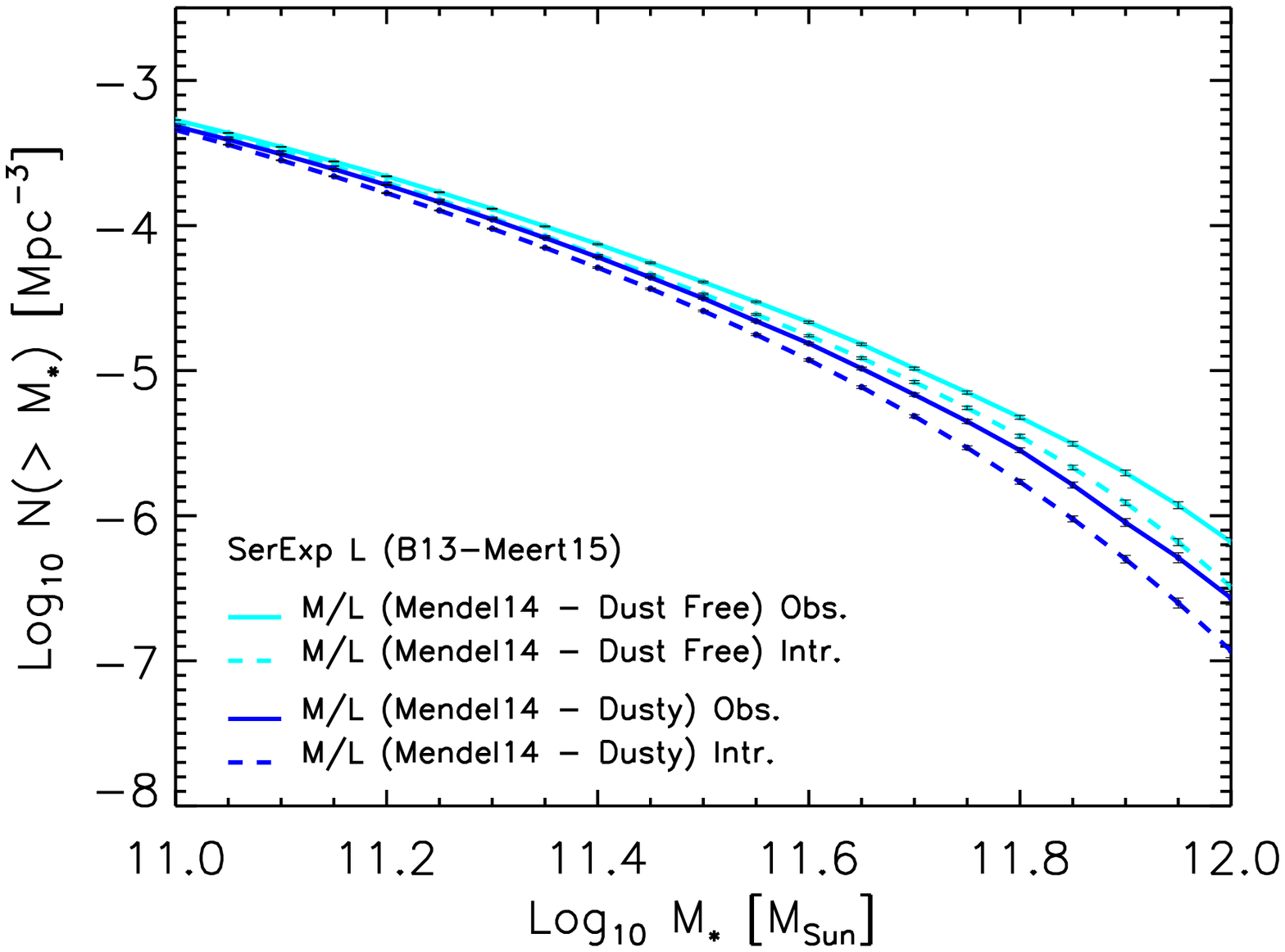}
 \includegraphics[width = 9.cm]{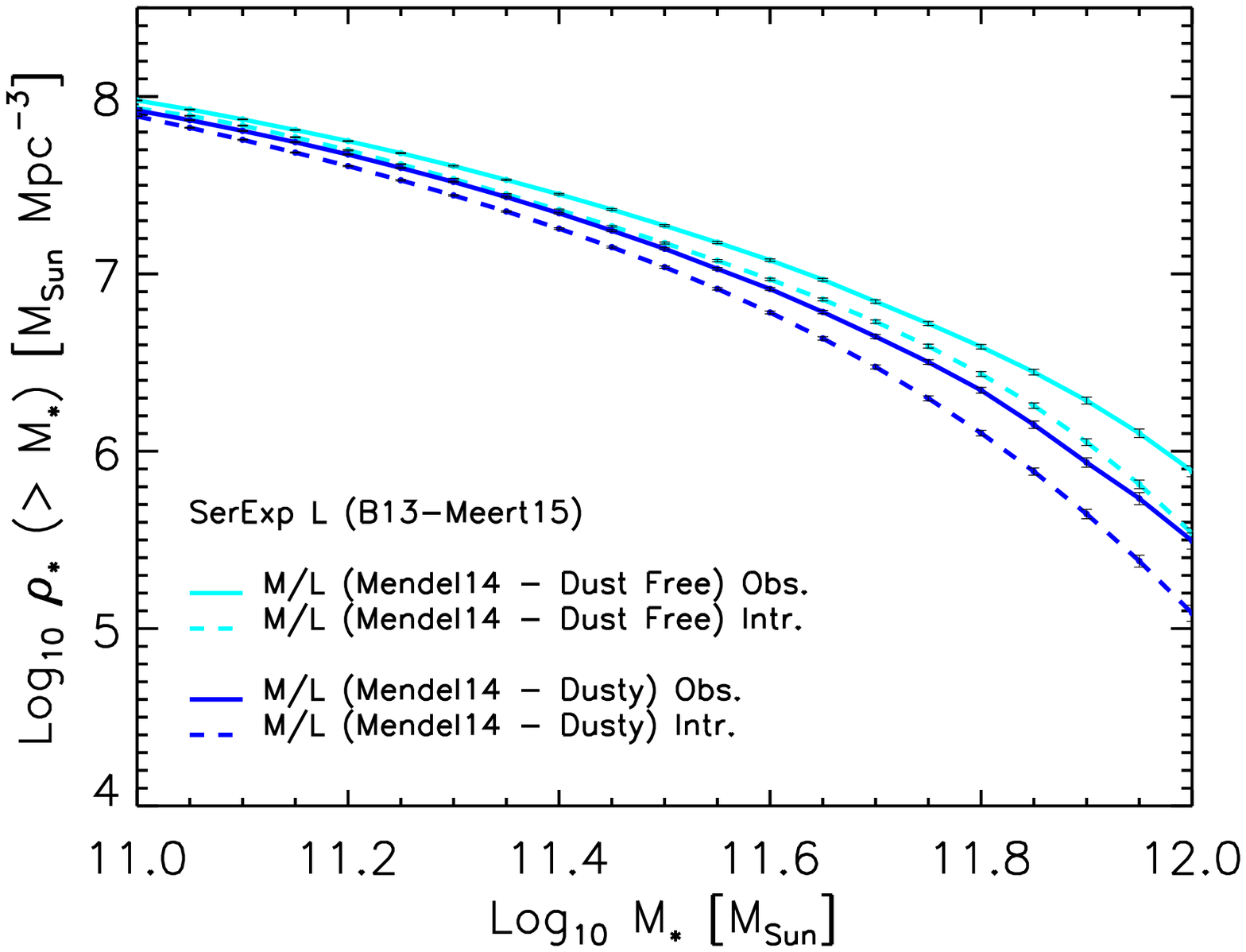}
 \caption{Cumulative number density (top) and stellar mass density (bottom) for the $M_*/L$ choices shown in Figure~\ref{errsSDSS}.  Each pair of solid and dashed curves shows the measured (error-broadened) and intrinsic counts for a given stellar population model (indicated by $M_*/L$).  Different sets of curves show different treatments of the stellar populations in massive galaxies; these can lead to differences as large as $\sim 0.3$~dex at $M_* \sim 10^{12}M_\odot$.  The curves in the top panel show our results in a format which is most directly relevant to models which match galaxies to dark matter halos. }
 \label{cumulative}
\end{figure}

Figure~\ref{errsSDSS} shows the measured (solid) and intrinsic (dashed) stellar mass functions for the dusty and dust-free models of Mendel14 (the reference photometry is SerExp of B13-Meert15, so comparing the solid curves here with those in Figure~\ref{compareMendel} shows that changing from Simard11 Sersic photometry to B13-Meert15 SerExp is small).  The dashed lines show the `error-corrected' measurements obtained using a modified $V^{-1}_{\rm max}$ method which accounts for errors (see Sheth 2017 for details).  We assumed measurement errors of 0.1~dex in all cases.  Clearly, measurement errors matter little below $10^{11}M_\odot$, but enhance $\phi(M_*)$ by more than 0.3~dex at $\sim 10^{12}M_\odot$, consistent with the analysis above.  Only at these rather large masses are the (statistical) measurement errors comparable to the systematic differences which arise from different treatments of the stellar population.  Note that this is of order the mass scale where the finite size of the survey volume is also beginning to matter.

\subsection{Cumulative counts and stellar mass density}
Having illustrated how systematics in $M_*/L$, $\ell$, and their associated measurement errors all impact the shape of the observed $\phi(M_*)$, we now present our results in two slightly different but common formats.  
The top panel of Figure~\ref{cumulative} shows the cumulative counts $\phi(\ge\! M_*)$, as this is directly relevant to Halo Model (Cooray \& Sheth 2002) based approaches which match galaxies to dark matter halos (e.g. Shankar et al. 2014).  The bottom panel shows $\rho_*(\ge\! M_*)$, the stellar mass density that is locked up in objects more massive than some limiting stellar mass $M_*$.  In both panels, we show the same models as in Figure~\ref{errsSDSS} (integrated to an upper limit of $10^{12.2}M_\odot$, beyond which discreteness effects matter).  Notice again that measurement errors matter little when the lower limit is $10^{11}M_\odot$, but can contribute 0.3~dex or more above $10^{12}M_\odot$.

\section{Comparison with recent work}\label{recent}

We now compare our findings with recent work.  

\begin{figure}
 \centering
 \includegraphics[width=9cm]{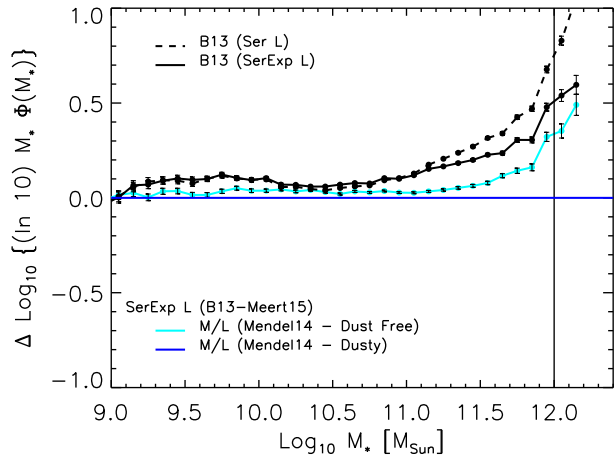}
 \caption{Comparison of $\phi(M_*)$ estimates based on dusty (solid blue) and dust-free (solid cyan) $M_*/L$ ratios of Mendel et al. (2014), and of B13 (solid black); all three curves used B13-Meert15 SerExp photometry.  Dashed-black curve represents B13 $M_*/L$ and B13-Meert15 Sersic photometry.  The $\sim 0.15$~dex difference between the solid cyan and black curves -- and not that between the blue and dashed-black curves -- is a better indicator of the true level of systematic difference between B13-Meert15 and Thanjavur et al. (2016) at high $M_*$.}
 \label{thanja}
\end{figure}

\subsection{Comparison with Thanjavur et al. (2016)}\label{Thanj}
Recently, Thanjavur et al. (2016) presented an analysis of $\phi(M_*)$ which was based on photometry of Simard11 along with the analysis of the $M_*/L$ values from Mendel et al. (2014).  Their Figure~7 shows that their `fiducial' estimate lies below the Sersic-based estimate of B13:  the offset is about 0.7~dex at $10^{12}M_\odot$.  However, we argued before that comparison with the B13 SerExp-based estimate would be more appropriate (Section~\ref{ell}).  We also noted that the difference will depend on whether $M_*/L$ assumed the galaxies were dusty or dust-free (Section~\ref{dust}).  Therefore, we have attempted to illustrate each of these effects as follows.  

\begin{figure*}
 \centering
 \includegraphics[width=14cm]{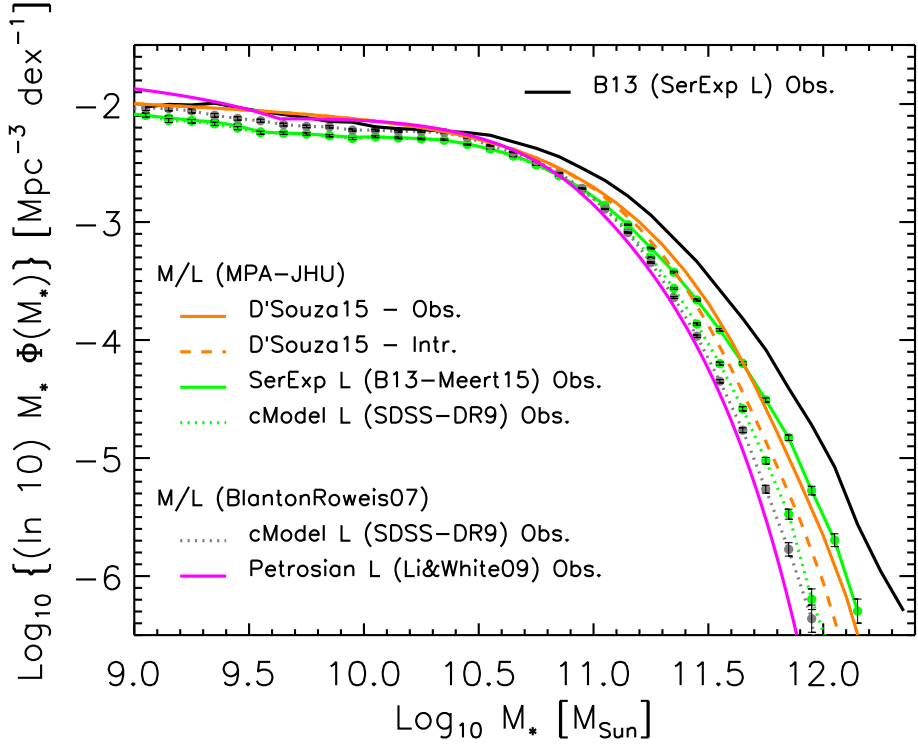}
 \caption{Comparison of a number of different determinations of $\phi(M_*)$ as labelled.  The curves of most interest here are the ones based on MPA-JHU $M_*/L$ (green lines), so differences between them are entirely due to how $\ell$ was estimated. The orange curves are the determination of $\phi(M_*)$ by D'Souza et al. (2015) obtained using the MPA-JHU $M_*/L$ values -- measured (solid) and intrinsic (dashed). The B13 SerExp estimate (black solid line), and the two Blanton-Roweis-based curves are shown as reference.}
 \label{phiMdsouza}
\end{figure*}

Our Figure~\ref{thanja} is intended to be similar to the bottom panel of Figure~7 in Thanjavur et al. (2016).  Although we do not have their `fiducial' $\ell$ values, we argued in Section~\ref{ell} that they must be close to the B13-Meert15 SerExp values.  Therefore, to define the fiducial model here, we use Mendel14 dusty $M_*/L$ and B13-Meert15 SerExp $\ell$ values.  The upper most (black dashed) curve shows the B13 $M_*/L$ and Sersic $\ell$ estimate.  The offset at $10^{12}M_\odot$ is indeed similar to that in Figure~7 of Thanjavur et al., suggesting that their `fiducial' photometry is indeed close to the B13-Meert15 SerExp. This offset is due in part to $M_*/L$ and in part to photometry.  Keeping $M_*/L$ the same as B13, but replacing Sersic with SerExp photometry, yields the solid black curve.  And keeping SerExp photometry, but replacing dusty with dust-free $M_*/L$ modifies the fiducial blue curve to the cyan one.  (We could also have said:  Keeping SerExp photometry, but replacing B13 $M_*/L$ for Mendel14 dust-free $M_*/L$ modifies the solid black curve to the cyan one.) The difference between this cyan curve and the black solid one represents the real level of systematic difference between B13-Meert15 and Thanjavur et al. (2016).  This difference is substantially smaller than 0.7~dex.  

\begin{figure}
 \centering
 \includegraphics[width=9cm]{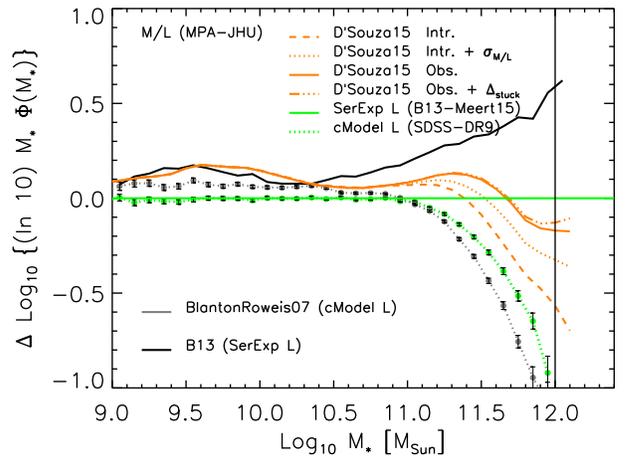}
 \caption{Similar to previous figure, but now shown in units of $\phi_{\rm fid}$ (solid green line) for which we combined the MPA-JHU $M_*/L$ ratios with the B13-Meert15 SerExp apparent brightnesses. The D'Souza et al. and B13 analyses agree, to within 0.1~dex, when the comparison is done using the same $M_*/L$.}
 \label{fixDsouza}
\end{figure}

\subsection{Comparison with D'Souza et al. (2015):  $M_*/L$}\label{DS}
We now return to the comparison with D'Souza et al. (2015).  
In Section~\ref{dsouza} we made the point that there was, in fact, little difference in photometry compared to B13-Meert15 SerExp.  So the following discussion addresses the question of $M_*/L$.

The lowest (magenta solid) curve in Figure~\ref{phiMdsouza} shows $\phi(M_*)$ reported by Li \& White (2009).  We only show it because it appears in D'Souza et al. (2015).  Li \& White used Petrosian luminosities (known even prior to 2009 to underestimate $\ell$ of massive galaxies) and $M_*/L$ ratios from Blanton \& Roweis (2007) (a choice Blanton-Roweis themselves cautioned against for the most massive galaxies; also see Figure~22 in Bernardi et al. 2010 and related discussion), so it is not surprising that it is the lowest.  Replacing Petrosian with {\tt cModel} DR7 magnitudes yields the dotted gray curve; it too lies substantially below all the others in the Figure.  Exchanging the Blanton-Roweis $M_*/L$ for MPA-JHU values increases the {\tt cModel} curve slightly more (dotted green curve), but the difference compared to the Blanton-Roweis curve (magenta) is more than 1~dex at $M_*\ge 10^{12}M_\odot$.

The black solid upper most curve comes from SerExp luminosities and the color-based prescription for $M_*/L$ of B13.  Using the MPA-JHU $M_*/L$ values  instead -- as this was the choice made by D'Souza et al. -- yields the curve labeled B13-Meert15 (solid green).  This green curve is in rather good agreement with the solid orange curve, which is our rendition of D'Souza et al.'s determination.  Since the only difference between the black and green curves is $M_*/L$, the agreement between the green and orange curves shows that, when the comparison is done using the same $M_*/L$, then D'Souza et al. (2015) are in good agreement with B13.  This is consistent with our finding (Figure~\ref{stacks2serexp}) that the median corrections to SDSS pipeline photometry were similar.

Strictly speaking, D'Souza et al. used a slightly different background cosmology ($\Omega_m=0.25$, $\Omega_\Lambda = 0.75$).  We expect this to cause negligible difference because, at $z\sim 0.15$, the luminosity distance in their cosmology is larger than in ours by a factor of 717.2/713.5, so their luminosities are about one percent brighter than ours.  
We checked this effect by scaling the B13 sample to D'Souza et al.'s cosmology and recomputing $\phi(M_*)$, finding it to be nearly indistinguishable from the original.  Hence, we expect this would also be true if we instead scaled D'Souza et al. to our cosmology.  Therefore, we have shown the determination of $\phi(M_*)$ by D'Souza et al. without applying any changes to their cosmology.

However, D'Souza et al. actually only report the intrinsic, error-corrected dashed orange curve (see their Table~2), so we have had to do some work to produce the error-broadened solid orange curve.  Each correction is small, so it is easier to see their effects on $\phi(M_*)$ when normalized by a fiducial value, which we take to be that for MPA-JHU $M_*/L$ and B13-Meert15 SerExp $\ell$ values (solid green line).  The steps we took are:

(i) The dashed orange curve in Figure~\ref{fixDsouza} shows D'Souza et al.'s intrinsic (error corrected) curve.  (All the orange curves in this figure were scaled to a Chabrier IMF by applying the 0.05~dex offset between the Kroupa and Chabrier IMFs.) 

(ii) The dotted orange curve shows the result of accounting for errors in the $M_*/L$ determination by using the 0.1~dex they report in our equation~(\ref{b10eq10}).  \\
The difference between the dotted orange curve and the solid black curve is the discrepancy with B13 which they report.  The difference in Figure~\ref{fixDsouza} is actually larger than that reported in Figure 7 of D'Souza et al. because they ignored the offset between the Kroupa and Chabrier IMFs.  While this is of order 0.3~dex at $10^{11.5}M_\odot$, the appropriate comparison is with the fiducial curve, for which the differences are larger than 0.3~dex only above $10^{12}M_\odot$.

(iii) There is another systematic difference between the D'Souza et al. photometry and the more traditional estimates of B13 (and Simard11).  For a given class of objects, D'Souza et al. use a single correction factor (based on their stacked images).  This does not include the fact that there is scatter around the median correction.  The B13 curves do include this scatter, so a fair comparison requires that we account for its effects.  In our discussion of Figure~\ref{stacks2serexp} we argued that most of the measured scatter is intrinsic -- there is no reason why the profile of each galaxy in the stack should be exactly the same -- so we have accounted for an additional $0.2/2.5$~dex correction to the orange dotted curve using our equation~(\ref{b10eq10}).  This produces the solid orange curve.  

(iv) Finally, contrary to what is written in the caption of Figure~4 of D'Souza et al., they actually used the mean instead of median correction.  The median is slightly larger -- using it instead yields the triple-dot-dashed line in Figure~\ref{fixDsouza}. This is a small effect, which is why, in Figure~\ref{phiMdsouza}, we treat the orange solid curve as the final corrected value.  \\
Clearly, once all of these small corrections have been made, the discrepancy with respect to B13-Meert SerExp is within about 0.1~dex (when using the same $M_*/L$).  This is much smaller than D'Souza et al. originally claimed was due to photometry.

Before moving on, we think it is important to note that below about $10^{10.5}M_\odot$ D'Souza et al. (2015) lie about 0.15~dex above the fiducial value, which is based on B13-Meert15 SerExp photometry.  We have argued that this photometry is likely very close to the `fiducial' photometry of Thanjavur et al. (2016).  Once differences in $M_*/L$ have been accounted for, the fiducial curve is similar to that of Thanjavur et al.: in fact, Figure~\ref{thanja} suggests that it is already slightly above the Thanjavur et al. determination, so D'Souza et al. would be even higher.  This offset is also present in their analysis of the luminosity function:  Figure~8 in D'Souza et al. (2015) shows that they lie about 0.2~dex above the $\phi(L)$ determination of B13.  One cannot appeal to $M_*/L$ differences to explain the offset in $\phi(L)$, so it must be due to other effects.  Since B13 were careful to establish that, at the faint end, their luminosity function was in excellent agreement with the $\phi(L)$ estimates of several other groups, the vertical offset between D'Souza et al. and B13 is surprising. The reason for this is an open question.

\subsection{Comparison with Moustakas et al. (2013) and Summary}

\begin{figure*}
 \centering
 \includegraphics[width = 14cm]{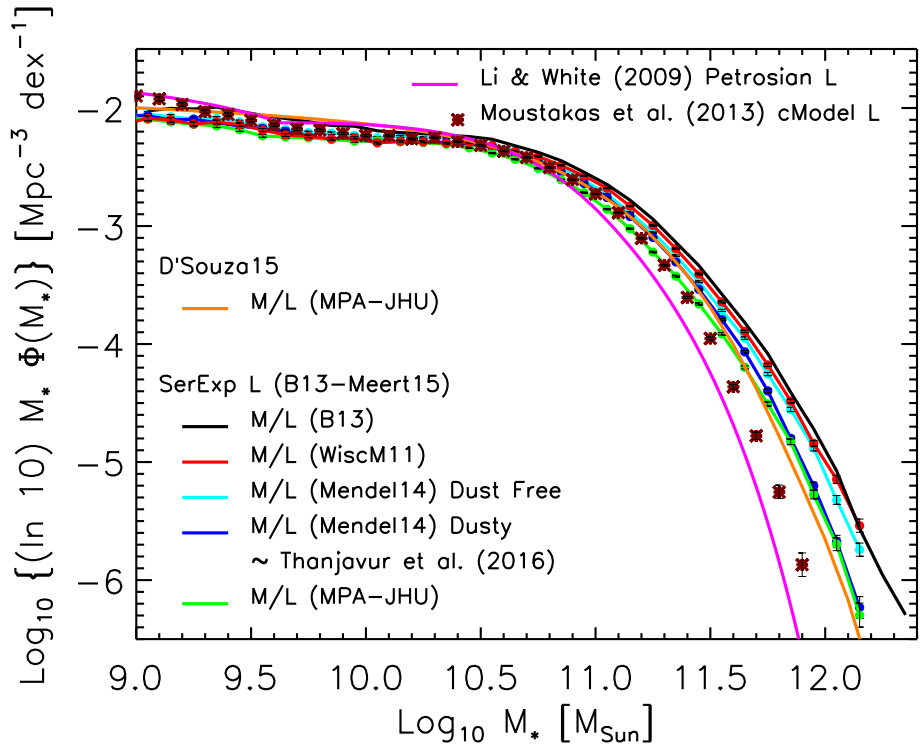}
 \caption{Representative $\phi(M_*)$ estimates before (Li \& White 2009; Moustakas et al. 2013) and after B13 as labeled. All curves and symbols only show error-broadened values.}
 \label{lit}
\end{figure*}

\begin{figure}
 \centering
 \includegraphics[width = 9cm]{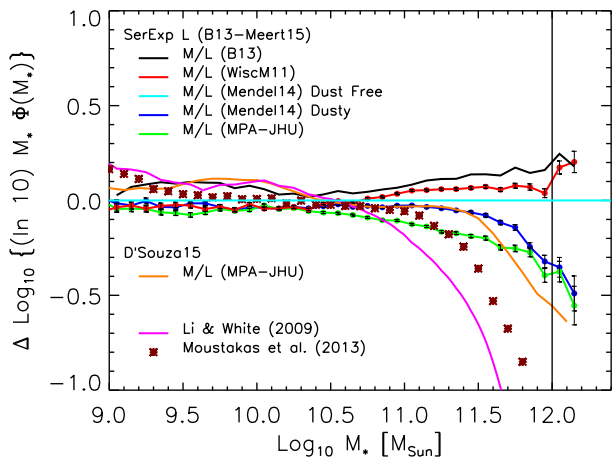}
 \caption{Similar to previous figure, but now shown in units of $\phi_{\rm fid}$ (solid cyan line) for which we combined the Mendel14 dust-free $M_*/L$ ratios with the B13-Meert15 SerExp apparent brightnesses.}
 \label{litResid}
\end{figure}

We have described how a variety of systematics affect determination of the stellar mass function.  To put these results in perspective, Figure~\ref{lit} shows the state of the art before and after B13.  Figure~\ref{litResid} shows the same curves normalized by the counts associated with the Mendel14 dust-free $M_*/L$ ratios and the B13-Meert15 SerExp apparent brightnesses (solid cyan curve).

The two bottom-most estimates are based on SDSS pipeline determinations of $\ell$.  The magenta solid curve shows the Li \& White (2009) estimate; we have already noted that it was biased low because of inappropriate choices for both photometry and stellar population modeling.  
The brown crosses show the estimate of Moustakas et al. (2013) which uses slightly better (but still SDSS pipeline) {\tt cModel} photometry $\ell$ and a different stellar population model (about which, more in Section~\ref{Moustakas}).  

The other curves are all based on more recent determinations of $\ell$ and they all result in more high-mass objects.  The solid orange and green curves show the result of using the same dusty and bursty MPA-JHU $M_*/L$ estimate, but with more sophisticated estimates of the apparent brightness (the stacking analysis of D'Souza et al. 2015 and SerExp values of B13-Meert15, respectively).  Both curves predict substantially more stellar mass than Moustakas et al. and Li \& White:  the increase is of order 1~dex and 2~dex at $10^{12}M_\odot$, respectively.  The small differences between these two curves (orange and green) show that photometry now contributes only of order 0.1~dex to the systematics error budget. That we are now discussing 0.1~dex systematics rather than 1 or 2~dex represents real progress in treatments of the photometry of the most massive galaxies.  

The other curves show results based on alternative treatments of the stellar populations.  The cyan and blue curves show the systematic effects of dust in the single burst Mendel14 models:  the cyan curves assume galaxies are dust-free, suggesting that, within the context of the same modelling and fitting framework, dust leads to differences up to 0.3~dex at $10^{12}M_\odot$.  However, there is no agreement even on the sign of the systematic.  The blue (Mendel14) and red (WiscM11) curves are both dusty models -- so differences between frameworks lead to $\ge 0.3$~dex differences already at $M_* \ge 10^{11.5}M_\odot$ -- but the dust-free version of the red curve would bring it down, close to the blue curve, rather than increase it.   Thus, we conclude that uncertainties in $M_*/L$ are at least 0.3~dex; Figure~\ref{litResid} in particular shows that they are substantially larger than photometry-related systematics.  

The upper most curve shows the estimate of B13.  Some of this can be traced to the fact that its calibration of $M_*/L$ (from color) is based on the {\tt Pegase} stellar population models, and these tend to produce the largest $\phi(M_*)$ estimates (see right hand panel of Figure~19 in Moustakas et al. 2013).  In this context, it is interesting to revisit the offset between Moustakas et al. and the others.  

\begin{figure*}
 \centering
 \includegraphics[width = 14cm]{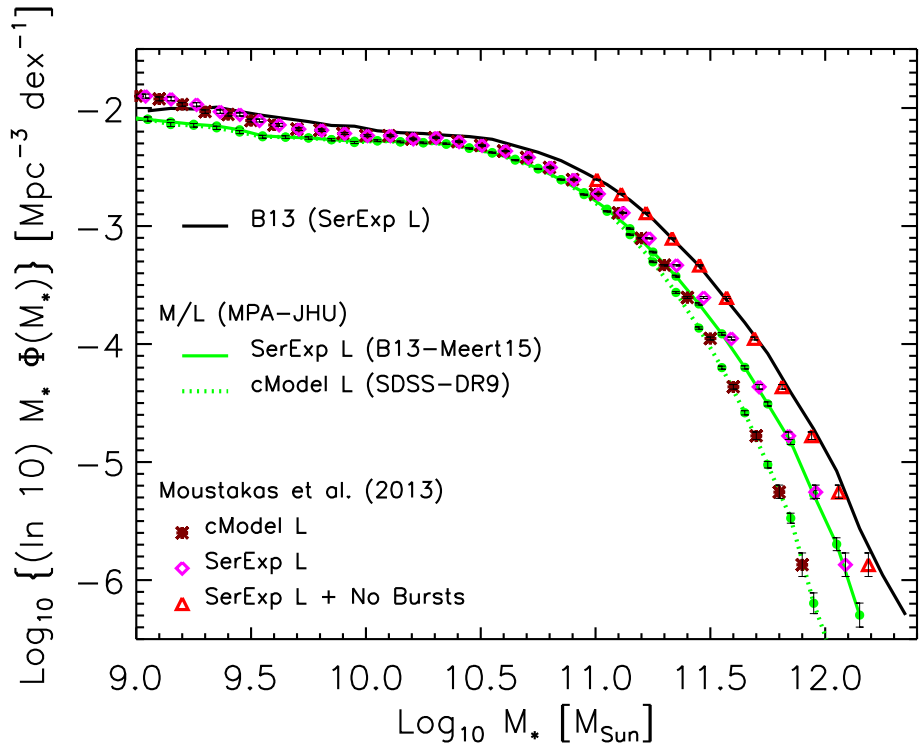}
 \caption{Effect on the $\phi(M_*)$ of Moustakas et al. (2013) when $\ell$ is increased from {\tt cModel} (brown crosses) to B13-Meert15 SerExp (magenta diamonds), and then when bursty models are not included when determining $M_*/L$ (red triangles).  Shifting from FSPS to {\tt Pegase} models yields an estimate which lies slightly above that of B13 (black solid curve). }
 \label{moustakas}
\end{figure*}

\subsubsection{Effects of bursts}\label{Moustakas}
We know that a systematic difference between the Moustakas et al. (2013) determination of $\phi(M_*)$ and that of Mendel et al. (2014) is due to $\ell$.  Another is due to the use of bursty (or not) models when determining $M_*/L$.  The left hand panel of Figure~19 in Moustakas et al. shows that fitting with single burst models would return $M_*/L$ values which are larger by at least 0.1~dex.  Mendel et al. (2014) report a similar difference, and go on to argue that single burst models are adequate for the vast majority of the galaxy population.  As this is even more likely to be true of the most massive galaxies, they do not consider bursty models at all.  Therefore, we now explore how the Moustakas et al $\phi(M_*)$ determination compares with the others once we have accounted for both $\ell$ and bursty-ness.  We do not have their $M_*$ determinations on an object-by-object basis, so we have transformed their $\phi(M_*)$ estimate in two steps as follows.  

The brown symbols in Figure~\ref{moustakas} show their fiducial determination.  It happens to be in good agreement with the dotted green curve.  This is a coincidence, because, although it too uses the same {\tt cModel} $\ell$, the $M_*/L$ determination is different (FSPS vs MPA-JHU, both bursty).  But we can use this to motivate a simple correction for the effect of $\ell$.  Namely, the solid green curve shows how $\phi(M_*)$ changes when the {\tt cModel} photometry is replaced with SerExp of B13-Meert15.  Therefore, we shift each of Moustakas' brown crosses in the $M_*$ direction by the same amount that the corresponding solid and dotted green curves differ.  The magenta diamonds show the result; since the brown crosses were similar to the dotted curve, these diamonds are similar to the solid curve.  

To remove the effect of bursts on their $M_*$ estimates, we add an additional 0.1~dex (the value suggested by their Figure~19) to their $M_*$ values:  this yields the red triangles.  These are now rather close to the determination of B13.  Accounting for the difference between the {\tt Pegase}-models (on which the B13 values were calibrated) and the FSPS models they used would shift the red triangles by a further 0.05~dex to the right.  We conclude that, once differences in $\ell$ and $M_*/L$ have been removed, Moustakas et al. (2013) are in good agreement with the others.  But we emphasize that failure to account for the differences in $\ell$ (at least) will lead to a severe underestimate of the true counts at large $M_*$.

\subsection{Cumulative counts}
\begin{figure*}
 \centering
 \includegraphics[width = 15cm]{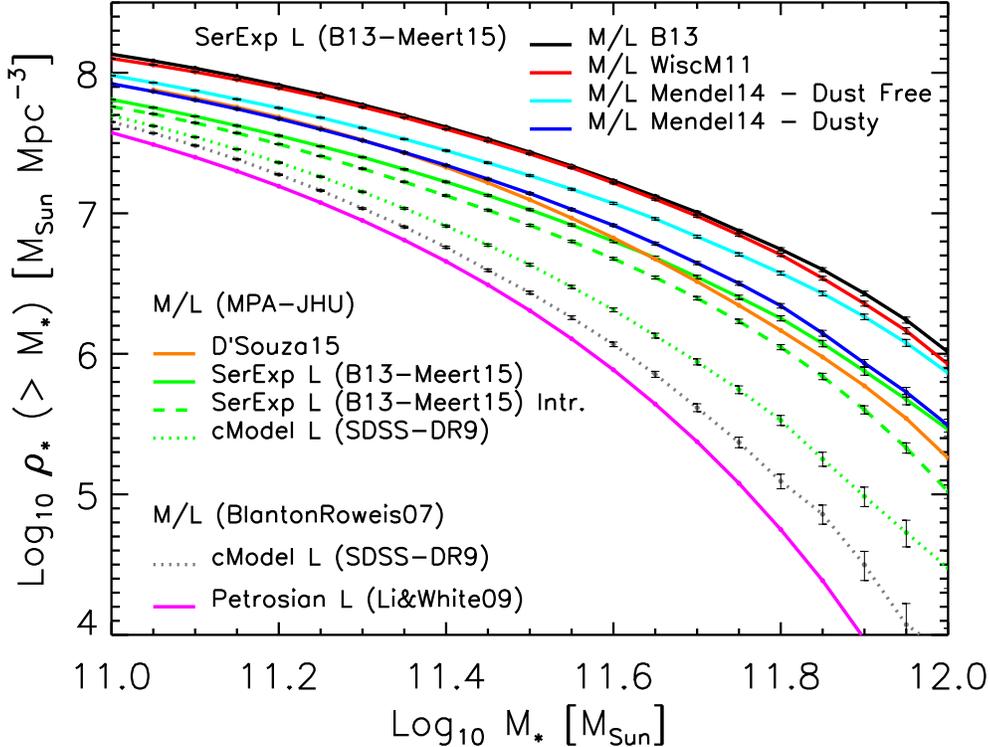}
 \caption{Cumulative stellar mass density for the curves shown in Figures~\ref{lit} and~\ref{moustakas}.  Solid and dashed green curves, which assume MPA-JHU $M_*/L$ values, show the observed error-broadened and intrinsic error-corrected estimates from B13-Meert15.  The other curves only show error-broadened values.  The D'Souza et al. (solid orange) and B13 (solid green) analyses agree, to within 0.1~dex, upto $10^{11.6}M_\odot$; both lie more than a factor of five above the estimates of Li \& White (2009).
At larger masses, while the top set of curves differ from one another by a factor of 2 at most (due to different assumptions in the star formation histories or/and presence of and effects of dust), they both lie about two orders of magnitude above the Li \& White estimate.}
 \label{dumulative}
\end{figure*}

Figure~\ref{dumulative} shows the cumulative stellar mass density for a selection of our stellar mass density $\rho_*(\ge\! M_*)$ determinations which summarize our findings. The bottom-most magenta solid curve shows the Li \& White (2009) estimate.  
The two dotted curves just above it show the result of using the most appropriate SDSS pipeline photometry available at the time (grey dotted), as well as a better estimate of $M_*/L$ which was also available (green dotted). These shifts alone account for more than 1~dex at $10^{12}M_\odot$.  Figure~\ref{lit} shows that the dotted green curve will also provide a good description of the results of Moustakas et al. (2013).  

The solid orange and green curves show the result of using the same dusty MPA-JHU $M_*/L$ estimate, but with more sophisticated estimates of the apparent brightness than the SDSS photometry (the stacking-based estimates of D'Souza et al. and SerExp fits of B13-Meert15, respectively).  Note that both curves predict substantially more stellar mass than Moustakas et al. and Li \& White:  the increase is of order 1~dex and 2~dex at $10^{12}M_\odot$, respectively.  The small differences between the orange and green curves show that photometry now contributes only of order 0.1~dex to the systematics error budget. 

The reasonable agreement between the blue curve and the orange and green ones is interesting.  Whereas the orange and green curves assumed bursty star formation histories, the blue curve did not.  If the single-burst assumption is more realistic at high masses, then we should shift the orange and green curves to the right by at least 0.1~dex (see discussion in previous section).  This would decrease the difference with respect to the blue curve.  While this is a coincidence, because they are based on different stellar population models (MPA-JHU and FSPS), it may indicate that consensus on this issue would significantly reduce systematics on $\phi(M_*)$. 

Unfortunately, this is not the full story.  The other solid curves show results based on other treatments of the stellar populations.  The cyan, blue and red curves correspond to the various  assumptions about dust discussed in Figure~\ref{litResid}.  

\begin{table*}
 \centering
 \caption{Table of observed (i.e. error-broadened) and intrinsic stellar mass functions $\Phi\equiv\log_{10}[\ln(10)M_*\phi(M_*)]$, in units of Mpc$^{-3}$dex$^{-1}$, as a function of ${\cal M}_*\equiv \log_{10}(M_*/M_\odot)$, for a range of stellar population models with SerExp photometry from Meert et al. (2015). M14$_{\rm d-f}$ and M14$_{\rm d}$ are the values associated to the dust-free and dusty estimate of $M_*$ from Mendel et al. (2014), respectively; and MPA and WM11 are the MPA-JHU and the WiscM11 (Chen et al. 2012) values associated to the $M_*$ estimates from tables {\tt galspecExtra} and {\tt stellarMassPCAWiscM11} available at {\tt www.sdss.org/dr12/spectro/galaxy}.  }
 \begin{tabular}{ccccccccccccc}
 \hline
  ${\cal M}_*$ & $\Phi^{\rm M14_{\rm d-f}}_{\rm Obs}$ & $\Phi^{\rm M14_{\rm d-f}}_{\rm Int}$ & $E^{\rm M14_{\rm d-f}}$ & $\Phi^{\rm M14_{\rm d}}_{\rm Obs}$ & $\Phi^{\rm M14_{\rm d}}_{\rm Int}$ & $E^{\rm M14_{\rm d}}$ & $\Phi^{\rm MPA}_{\rm Obs}$ & $\Phi^{\rm MPA}_{\rm Int}$ & $E^{\rm MPA}$ & $\Phi^{\rm WM11}_{\rm Obs}$ & $\Phi^{\rm WM11}_{\rm Int}$ & $E^{\rm WM11}$\\
 \hline
$ 9.05$ & $-2.051$ & $-2.069$ & $ 0.022$ & $-2.066$ & $-2.088$ & $ 0.019$ & $-2.097$ & $-2.103$ & $ 0.018$ & $-2.088$ & $-2.104$ & $ 0.021$ \\
$ 9.15$ & $-2.075$ & $-2.080$ & $ 0.020$ & $-2.101$ & $-2.095$ & $ 0.017$ & $-2.117$ & $-2.124$ & $ 0.017$ & $-2.101$ & $-2.112$ & $ 0.019$ \\
$ 9.25$ & $-2.092$ & $-2.091$ & $ 0.018$ & $-2.095$ & $-2.108$ & $ 0.016$ & $-2.138$ & $-2.148$ & $ 0.015$ & $-2.136$ & $-2.122$ & $ 0.017$ \\
$ 9.35$ & $-2.089$ & $-2.106$ & $ 0.016$ & $-2.124$ & $-2.138$ & $ 0.014$ & $-2.152$ & $-2.170$ & $ 0.014$ & $-2.118$ & $-2.136$ & $ 0.016$ \\
$ 9.45$ & $-2.118$ & $-2.139$ & $ 0.015$ & $-2.154$ & $-2.163$ & $ 0.013$ & $-2.185$ & $-2.207$ & $ 0.013$ & $-2.141$ & $-2.164$ & $ 0.014$ \\
$ 9.55$ & $-2.159$ & $-2.167$ & $ 0.014$ & $-2.175$ & $-2.188$ & $ 0.012$ & $-2.236$ & $-2.237$ & $ 0.012$ & $-2.187$ & $-2.206$ & $ 0.013$ \\
$ 9.65$ & $-2.183$ & $-2.196$ & $ 0.012$ & $-2.198$ & $-2.221$ & $ 0.011$ & $-2.246$ & $-2.252$ & $ 0.011$ & $-2.229$ & $-2.236$ & $ 0.012$ \\
$ 9.75$ & $-2.208$ & $-2.212$ & $ 0.011$ & $-2.242$ & $-2.253$ & $ 0.010$ & $-2.253$ & $-2.261$ & $ 0.010$ & $-2.241$ & $-2.252$ & $ 0.011$ \\
$ 9.85$ & $-2.213$ & $-2.223$ & $ 0.010$ & $-2.263$ & $-2.265$ & $ 0.010$ & $-2.267$ & $-2.273$ & $ 0.009$ & $-2.262$ & $-2.257$ & $ 0.010$ \\
$ 9.95$ & $-2.231$ & $-2.239$ & $ 0.010$ & $-2.269$ & $-2.279$ & $ 0.009$ & $-2.281$ & $-2.280$ & $ 0.008$ & $-2.252$ & $-2.274$ & $ 0.009$ \\
$10.05$ & $-2.252$ & $-2.246$ & $ 0.009$ & $-2.290$ & $-2.286$ & $ 0.008$ & $-2.279$ & $-2.283$ & $ 0.008$ & $-2.296$ & $-2.285$ & $ 0.009$ \\
$10.15$ & $-2.240$ & $-2.244$ & $ 0.008$ & $-2.285$ & $-2.285$ & $ 0.008$ & $-2.286$ & $-2.292$ & $ 0.007$ & $-2.282$ & $-2.287$ & $ 0.008$ \\
$10.25$ & $-2.252$ & $-2.252$ & $ 0.007$ & $-2.286$ & $-2.291$ & $ 0.007$ & $-2.293$ & $-2.300$ & $ 0.006$ & $-2.287$ & $-2.293$ & $ 0.007$ \\
$10.35$ & $-2.250$ & $-2.265$ & $ 0.006$ & $-2.292$ & $-2.298$ & $ 0.006$ & $-2.307$ & $-2.323$ & $ 0.006$ & $-2.298$ & $-2.294$ & $ 0.007$ \\
$10.45$ & $-2.274$ & $-2.295$ & $ 0.006$ & $-2.304$ & $-2.321$ & $ 0.006$ & $-2.338$ & $-2.361$ & $ 0.006$ & $-2.291$ & $-2.308$ & $ 0.006$ \\
$10.55$ & $-2.314$ & $-2.337$ & $ 0.005$ & $-2.335$ & $-2.365$ & $ 0.005$ & $-2.382$ & $-2.409$ & $ 0.005$ & $-2.318$ & $-2.344$ & $ 0.005$ \\
$10.65$ & $-2.357$ & $-2.392$ & $ 0.005$ & $-2.391$ & $-2.421$ & $ 0.005$ & $-2.434$ & $-2.476$ & $ 0.005$ & $-2.370$ & $-2.390$ & $ 0.005$ \\
$10.75$ & $-2.421$ & $-2.465$ & $ 0.005$ & $-2.449$ & $-2.497$ & $ 0.005$ & $-2.511$ & $-2.563$ & $ 0.005$ & $-2.411$ & $-2.451$ & $ 0.005$ \\
$10.85$ & $-2.504$ & $-2.561$ & $ 0.005$ & $-2.540$ & $-2.592$ & $ 0.005$ & $-2.608$ & $-2.664$ & $ 0.005$ & $-2.486$ & $-2.533$ & $ 0.005$ \\
$10.95$ & $-2.611$ & $-2.675$ & $ 0.005$ & $-2.639$ & $-2.702$ & $ 0.005$ & $-2.717$ & $-2.790$ & $ 0.005$ & $-2.576$ & $-2.632$ & $ 0.005$ \\
$11.05$ & $-2.732$ & $-2.812$ & $ 0.005$ & $-2.758$ & $-2.844$ & $ 0.005$ & $-2.857$ & $-2.945$ & $ 0.005$ & $-2.680$ & $-2.760$ & $ 0.005$ \\
$11.15$ & $-2.885$ & $-2.975$ & $ 0.005$ & $-2.919$ & $-3.011$ & $ 0.005$ & $-3.022$ & $-3.127$ & $ 0.006$ & $-2.831$ & $-2.918$ & $ 0.005$ \\
$11.25$ & $-3.055$ & $-3.159$ & $ 0.006$ & $-3.097$ & $-3.207$ & $ 0.006$ & $-3.220$ & $-3.329$ & $ 0.006$ & $-2.996$ & $-3.100$ & $ 0.005$ \\
$11.35$ & $-3.252$ & $-3.369$ & $ 0.007$ & $-3.304$ & $-3.427$ & $ 0.007$ & $-3.424$ & $-3.551$ & $ 0.007$ & $-3.191$ & $-3.303$ & $ 0.006$ \\
$11.45$ & $-3.472$ & $-3.601$ & $ 0.008$ & $-3.536$ & $-3.673$ & $ 0.008$ & $-3.660$ & $-3.795$ & $ 0.009$ & $-3.406$ & $-3.531$ & $ 0.007$ \\
$11.55$ & $-3.715$ & $-3.841$ & $ 0.009$ & $-3.793$ & $-3.942$ & $ 0.009$ & $-3.913$ & $-4.067$ & $ 0.011$ & $-3.642$ & $-3.772$ & $ 0.008$ \\
$11.65$ & $-3.950$ & $-4.114$ & $ 0.011$ & $-4.066$ & $-4.249$ & $ 0.012$ & $-4.198$ & $-4.360$ & $ 0.013$ & $-3.892$ & $-4.043$ & $ 0.010$ \\
$11.75$ & $-4.254$ & $-4.410$ & $ 0.014$ & $-4.397$ & $-4.616$ & $ 0.015$ & $-4.507$ & $-4.681$ & $ 0.017$ & $-4.176$ & $-4.344$ & $ 0.013$ \\
$11.85$ & $-4.554$ & $-4.732$ & $ 0.018$ & $-4.799$ & $-5.010$ & $ 0.022$ & $-4.828$ & $-5.079$ & $ 0.023$ & $-4.483$ & $-4.673$ & $ 0.017$ \\
$11.95$ & $-4.881$ & $-5.121$ & $ 0.024$ & $-5.202$ & $-5.457$ & $ 0.033$ & $-5.276$ & $-5.500$ & $ 0.036$ & $-4.840$ & $-4.997$ & $ 0.023$ \\
$12.05$ & $-5.321$ & $-5.548$ & $ 0.037$ & $-5.673$ & $-5.984$ & $ 0.052$ & $-5.696$ & $-6.073$ & $ 0.054$ & $-5.146$ & $-5.377$ & $ 0.034$ \\
$12.15$ & $-5.742$ & $-6.039$ & $ 0.055$ & $-6.233$ & $-6.558$ & $ 0.092$ & $-6.297$ & $-6.785$ & $ 0.102$ & $-5.539$ & $-5.900$ & $ 0.056$ \\
 \hline
 \end{tabular}
 \label{phiMtable}
\end{table*}

Whereas all these differences are systematic, there is also a statistical effect associated with measurement errors, which broaden the measured curve relative to its intrinsic value.  Comparison of the solid and dashed green curves shows the impact of 0.1~dex statistical measurement errors. Applying this difference to the other curves is a reasonable way of quantifying the effect of measurement errors on the other $\phi(M_*)$ determinations.  

\section{Conclusions}
When comparing different determinations of $\phi(M_*)$, it is important to separate effects which are due to differences in estimating the apparent brightness from those which arise from modeling the stellar population:  what we refered to as $\ell$ and $M_*/L$.  If the algorithm for assigning $M_*/L$ is the same, then different estimates of $\ell$ appear to give rise to order 1~dex differences in $\phi(M_*)$ at $10^{12}M_\odot$ (Figure~\ref{residLM}).  However, much of this is driven by the discrepancy between SDSS pipeline photometry and more recent work based on Sersic or SerExp fits.  We showed that these recent analyses agree, to within 0.1~dex, upto $\sim 10^{12}M_\odot$ (or higher).  Recent claims of larger discrepancies in the literature are primarily due to $M_*/L$ differences.  

Once differences in $M_*/L$ have been accounted for, the `fiducial' $\phi(M_*)$ estimate of Thanjavur et al. (2016) is in good agreement with that based on SerExp photometry of Bernardi et al. (2013) and Meert et al. (2015) (Figure~\ref{thanja}).  This implies that the `fiducial' photometry of Thanjavur et al. -- which is based on the work of Simard et al. (2011) -- is in good agreement with the B13-Meert15 SerExp photometry.  

In addition, we showed that the median corrections applied to the SDSS pipeline photometry by D'Souza et al. (2015) are essentially the same as those advocated by B13 (Figure~\ref{stacks2serexp}).  Comparison of the associated $\phi(M_*)$ is more complicated, because there is substantial scatter around this median ($\ge 0.15$~mags), most of which is intrinsic (c.f. Figure~\ref{stacks2serexp} and related discussion).  This matters most at the luminous end of $\phi(L)$ and hence the high-mass end of $\phi(M_*)$.  Object-by-object analyses like B13-Meert15 include this effect trivially, but stacking-based analyses (like D'Souza et al.) cannot.  Adding the effect of this scatter to the D'Souza et al. $\phi(M_*)$ estimate brings it, too, to within 0.1~dex of the B13-Meert15 SerExp-based estimate (Figure~\ref{fixDsouza}).

Thus, all three groups agree that, if the same $M_*/L$ algorithm is used, then the stellar mass density in objects more massive than $10^{11.3}M_\odot$ is at least $5\times$ larger than estimates based on SDSS pipeline photometry (Figure~\ref{dumulative}), as first noted by B13.  That we are now discussing 0.1~dex differences between groups, rather than factors of 5, represents substantial recent progress in the photometry of massive galaxies.  

When photometric parameters are held fixed, differences between stellar population treatments result in $\phi(M_*)$ estimates which can differ by more than a factor of three at $10^{11.3} - 10^{12}M_\odot$ (Figures~\ref{compareML}, \ref{residML}, \ref{lit} and~\ref{litResid}).  These differences arise because of differing assumptions about presence of and effects of dust (Figures~\ref{compareMendel} and~\ref{dustNoDust}) and of whether or not the star formation histories were bursty (Figure~\ref{moustakas}).  Hence, it is systematics in modeling the stellar population, and not the photometry, which now dominate the $\phi(M_*)$ error budget.  For example, there is currently not even agreement on whether allowing for dust should increase or decrease the estimated $M_*/L$ (Appendix~A).  

The shape of $\phi(M_*)$ means that errors -- statistical or systematic -- mainly matter at the massive end (equation~\ref{b10eq10}).  This means that there may be general agreement on $\phi(M_*)$ at masses below $\sim 10^{10.5}M_\odot$ (Figure~\ref{residML}), even though there is substantial disagreement about $M_*/L$ (Appendix~A). Therefore, the 0.1~dex agreement between various determinations of $\phi(M_*)$ at lower masses should not be used to argue that $M_*/L$ in low mass galaxies is well understood!  At higher masses, stastistical measurement errors on the value of $M_*$ can have a big impact:  0.1~dex errors give rise to 0.3~dex differences in $\phi(M_*)$ above $10^{12}M_\odot$ (Figures~\ref{cumulative} and~\ref{dumulative}).  As Shankar et al. (2014) have emphasized, these differences matter greatly for matching galaxies to dark matter halos, and currently limit what we can learn about galaxy formation from this match.

While we do not claim that one particular stellar population model is better than another -- Table~\ref{phiMtable} provides our results in tabular form for a variety of such models (for the photometry, we use the SerExp magnitudes from Meert et al. 2015) --  we hope that our summary of the current state of the field will spur work towards reducing these systematic differences.  For example, we have focussed on the systematics in determining $\phi(M_*)$.  However, the clustering of these objects also depends on $M_*$, especially at the high masses of interest here.  Therefore, it may be that Halo Model interpretations of the dependence of clustering on $M_*$ can be used to help decide between different stellar population assumptions.  Until such systematics have been reduced, using the same models and photometry whenever low and high redshift samples are compared -- as was done by Bernardi et al. (2016) when comparing the CMASS galaxies at $z\sim 0.6$ to this SDSS sample at $z\sim 0.1$ -- is essential.  The statistical power of large data sets is now sufficiently large that being careless about this will lead to systematically biased conclusions.

\subsection*{Acknowledgements}
We are grateful to the staff of the LYTE center for its hospitality when this work was completed.

\appendix
\section{Dependence of $M_*/L$ values on stellar population modeling and dust}

The main text made the point that it is important to separate the effects of $\ell$ from those of $M_*/L$ on $\phi(M_*)$.  This Appendix compares the $M_*/L$ values of the different groups, and shows how the inclusion of dust impacts the estimates.  In all cases, the photometry is SerExp of B13-Meert15, and the IMF is Chabrier.  

\begin{figure}
 \centering
 \includegraphics[width = 9cm]{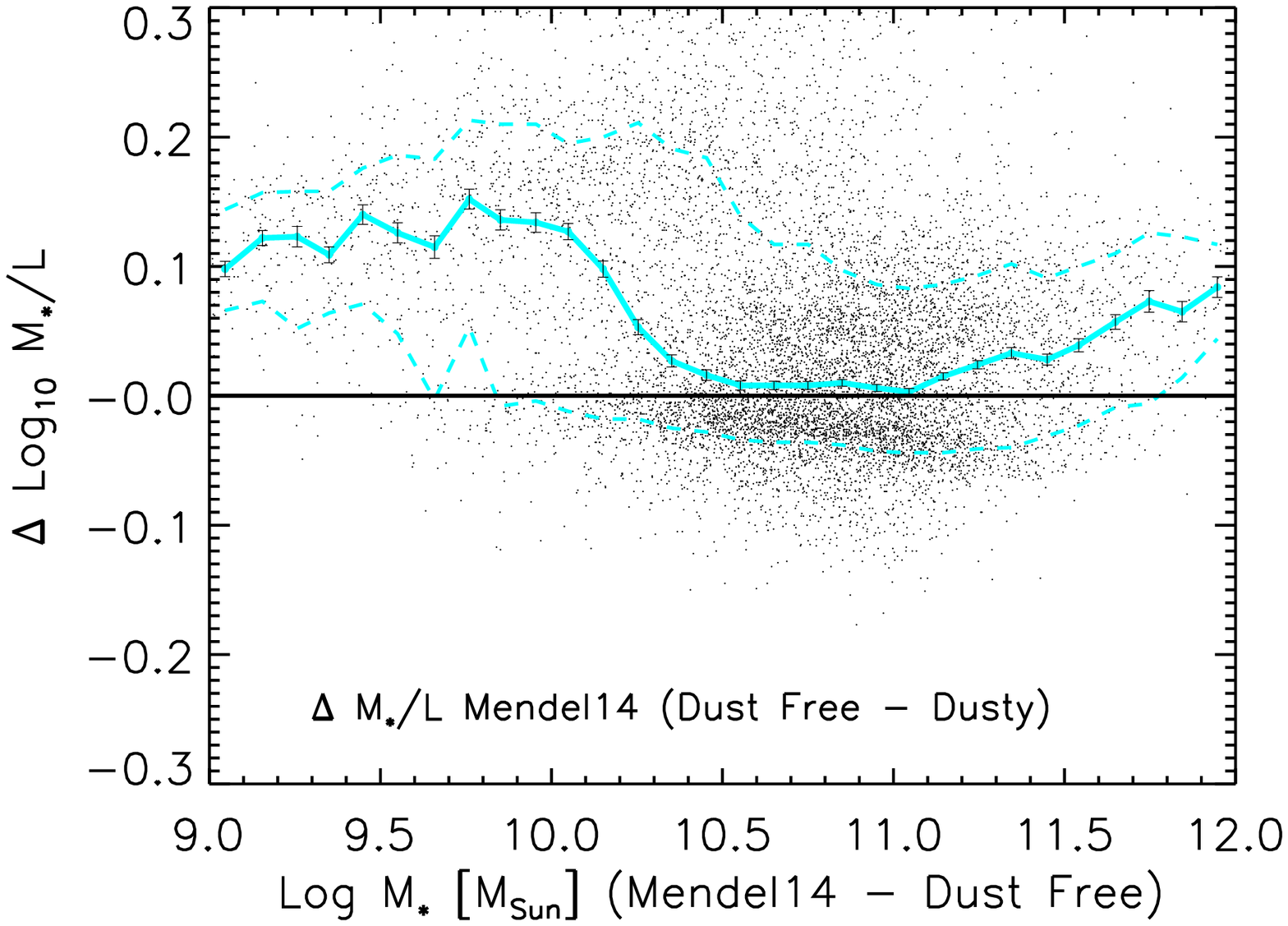}
 \caption{Comparison of the dust-free and dusty $M_*/L$ values from Mendel et al. (2014).  At the highest masses, the dust-free models return higher masses.  Although we only show $10^4$ objects, the solid line shows the median relation for the full sample, and dotted lines show the range within which 68\% of the full sample lies.}
 \label{mendelDust}
\end{figure}

\begin{figure}
 \centering
 \includegraphics[width = 9cm]{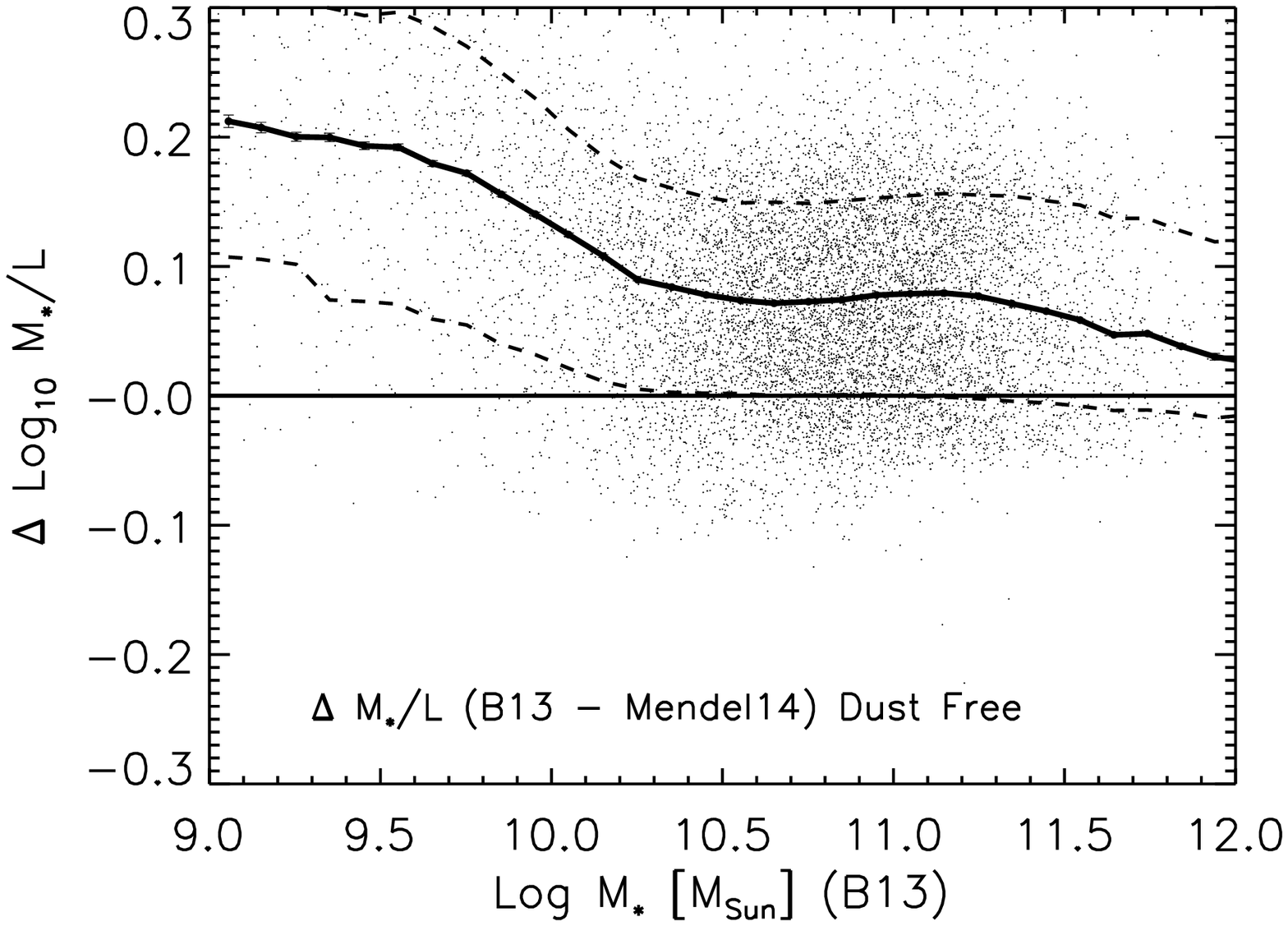}
 \caption{Comparison of the B13 and Mendel et al. (2014) dust-free $M_*/L$ values for a Chabrier IMF. Symbols and line styles same as in Figure~\ref{mendelDust}.}
 \label{mendelB13}
\end{figure}

\begin{figure}
 \centering
 \includegraphics[width = 9cm]{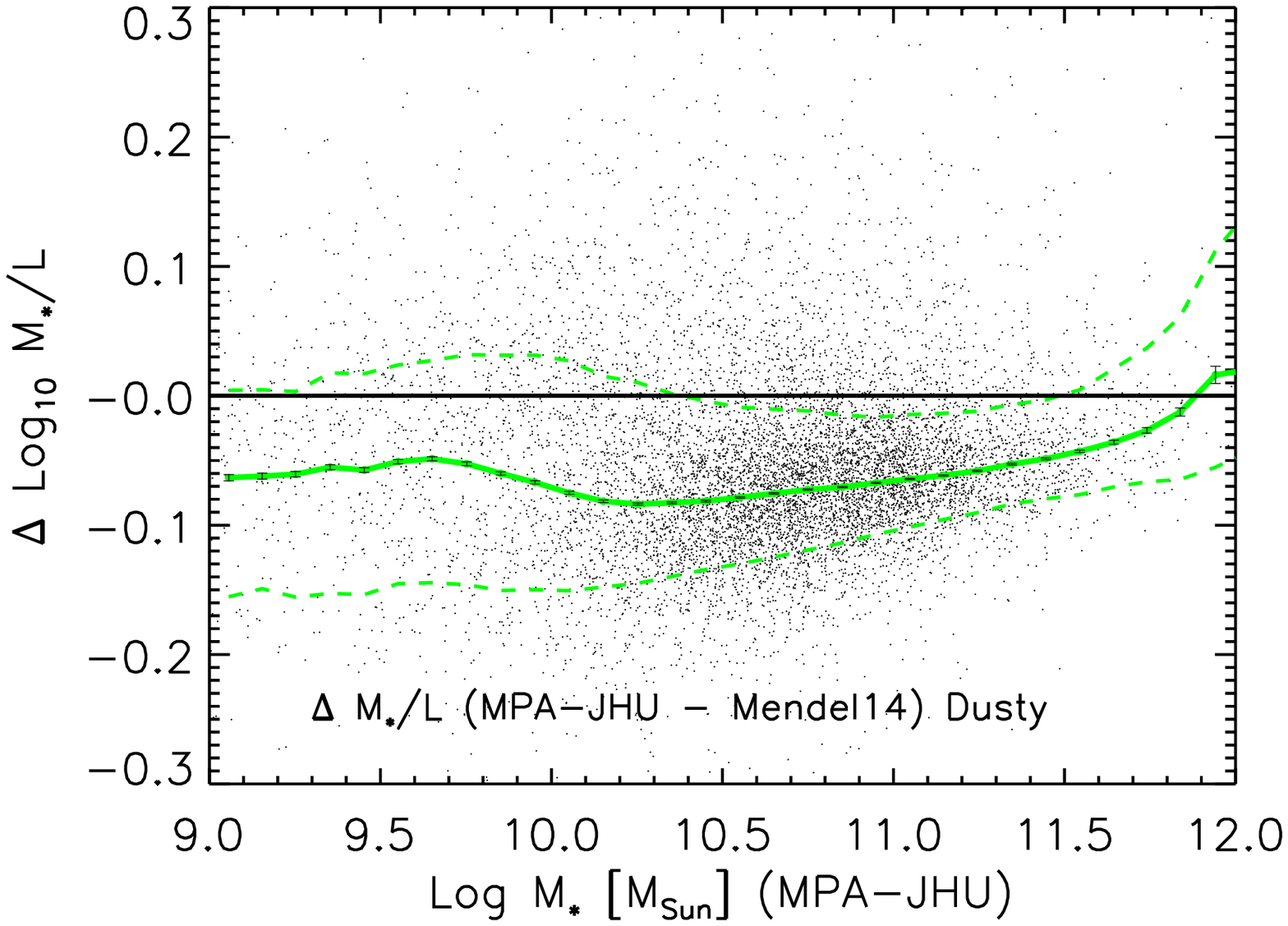}
 \caption{Comparison of the MPA-JHU and Mendel et al. (2014) $M_*/L$ values for a Chabrier IMF.  (Both estimates assume galaxies are dusty. However, the MPA-JHU includes bursty models while Mendel et al. do not.) Symbols and line styles same as in Figure~\ref{mendelDust}.}
 \label{mendelJHU}
\end{figure}

\begin{figure}
 \centering
 \includegraphics[width = 9cm]{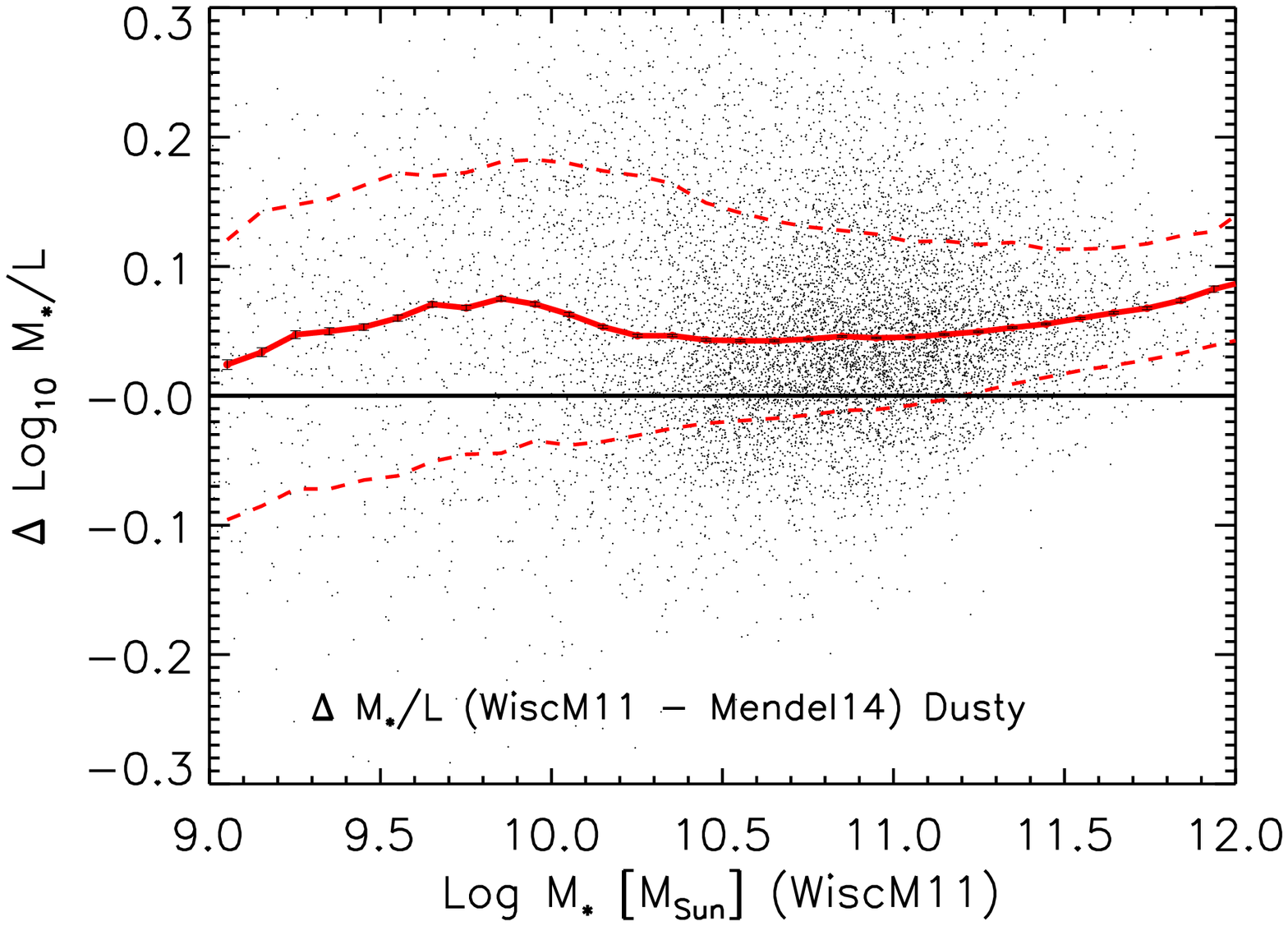}
 \caption{Comparison of the Wisconsin and Mendel et al. (2014) $M_*/L$ values for a Chabrier IMF.  (Both estimates assume galaxies are dusty.)  Comparison with the previous figure suggests that the Wisconsin $M_*/L$ values are about 0.1~dex larger than the MPA-JHU values. Symbols and line styles same as in Figure~\ref{mendelDust}.}
 \label{mendelWisc}
\end{figure}

We begin with a comparison of the dusty and dust-free $M_*/L$ values from Mendel et al. (2014) in Figure~\ref{mendelDust}. The solid cyan line shows the median difference for a number of narrow bins in $M_*$.  The dashed lines show the range around this median which contains 68\% of the objects in each $M_*$ bin.  The (large!) differences at low $M_*$ are easy to understand:  if dust is not accounted-for red spirals will be assigned larger masses.  However, these are the objects of least relevance to our work.  At intermediate masses, there appear to be two populations, one of which is assigned smaller masses when moving from dusty to dust-free models.  This is mostly irrelevant for the stellar mass function, since it is relatively flat at masses below about $10^{11}M_\odot$, so shifts in mass make little difference.  At the higher masses of most interest to us, these shifts do matter, but in this regime the dust-free models again return higher masses.  

\begin{figure}
 \centering
 \includegraphics[width = 9cm]{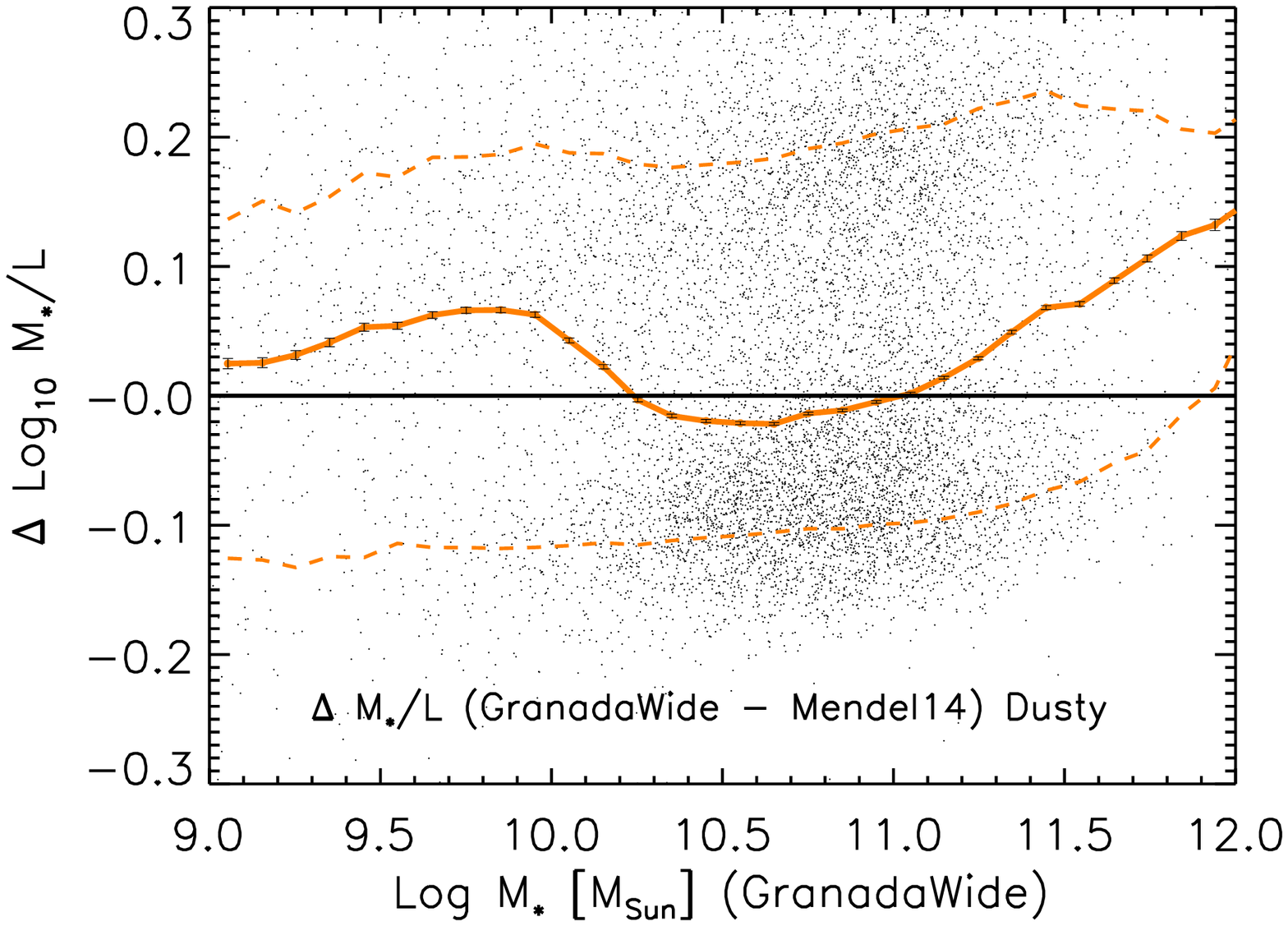}
 \includegraphics[width = 9cm]{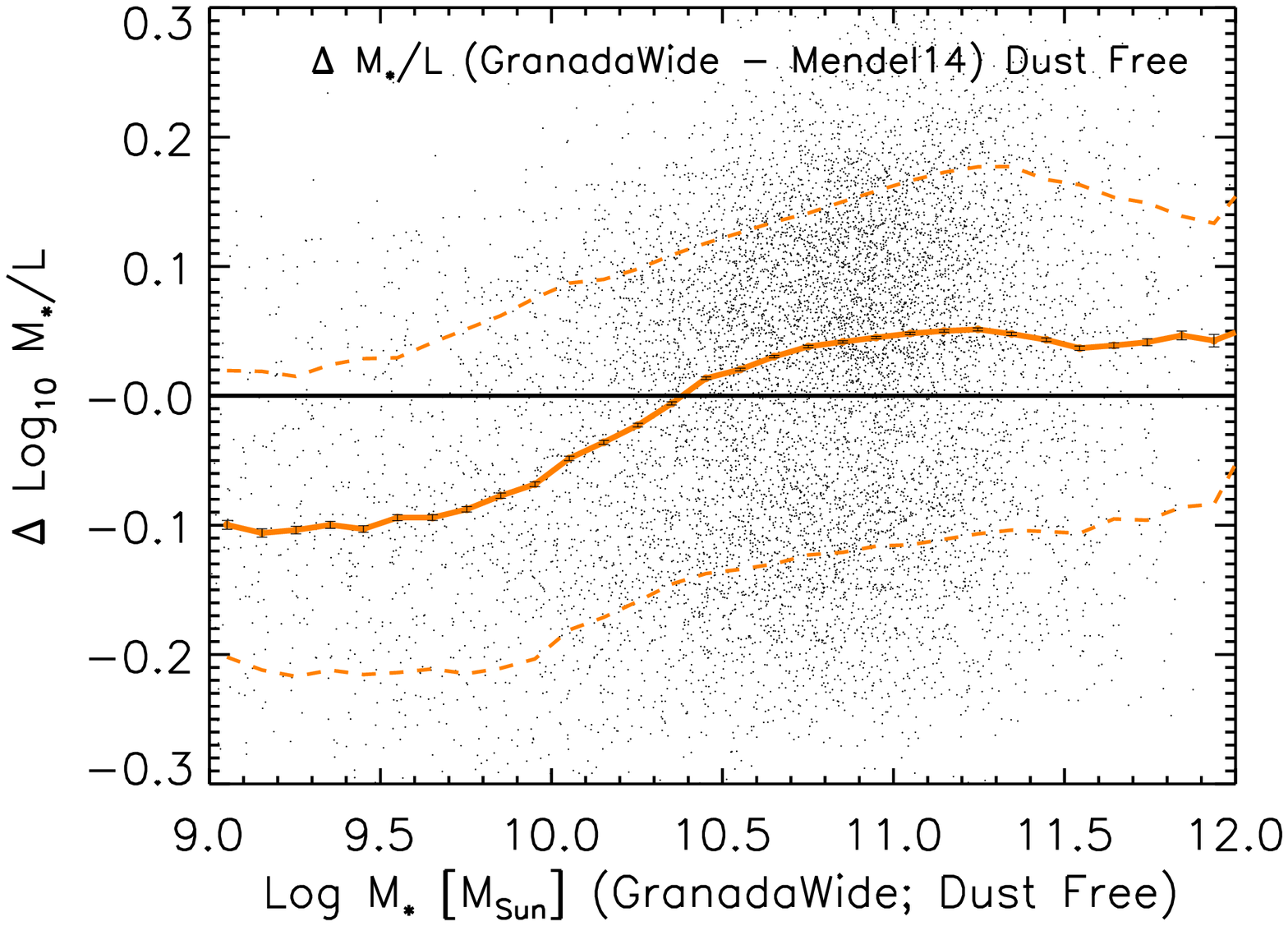}
 \caption{Comparison of the GranadaWide and Mendel et al. (2014) $M_*/L$ values for a Chabrier IMF. Symbols and line styles same as in Figure~\ref{mendelDust}. Top and bottom panels show results when galaxies are assumed to be dusty and dust free, respectively.  In contrast to the previous two figures, there appear to be two populations of galaxies which are approximately equally populated in the dust free models, but not so when dust is included.  As a result, the scatter around the median is much larger here than in the previous two figures.  }
 \label{mendelGranada}
\end{figure}

\begin{figure}
 \centering
 \includegraphics[width = 9cm]{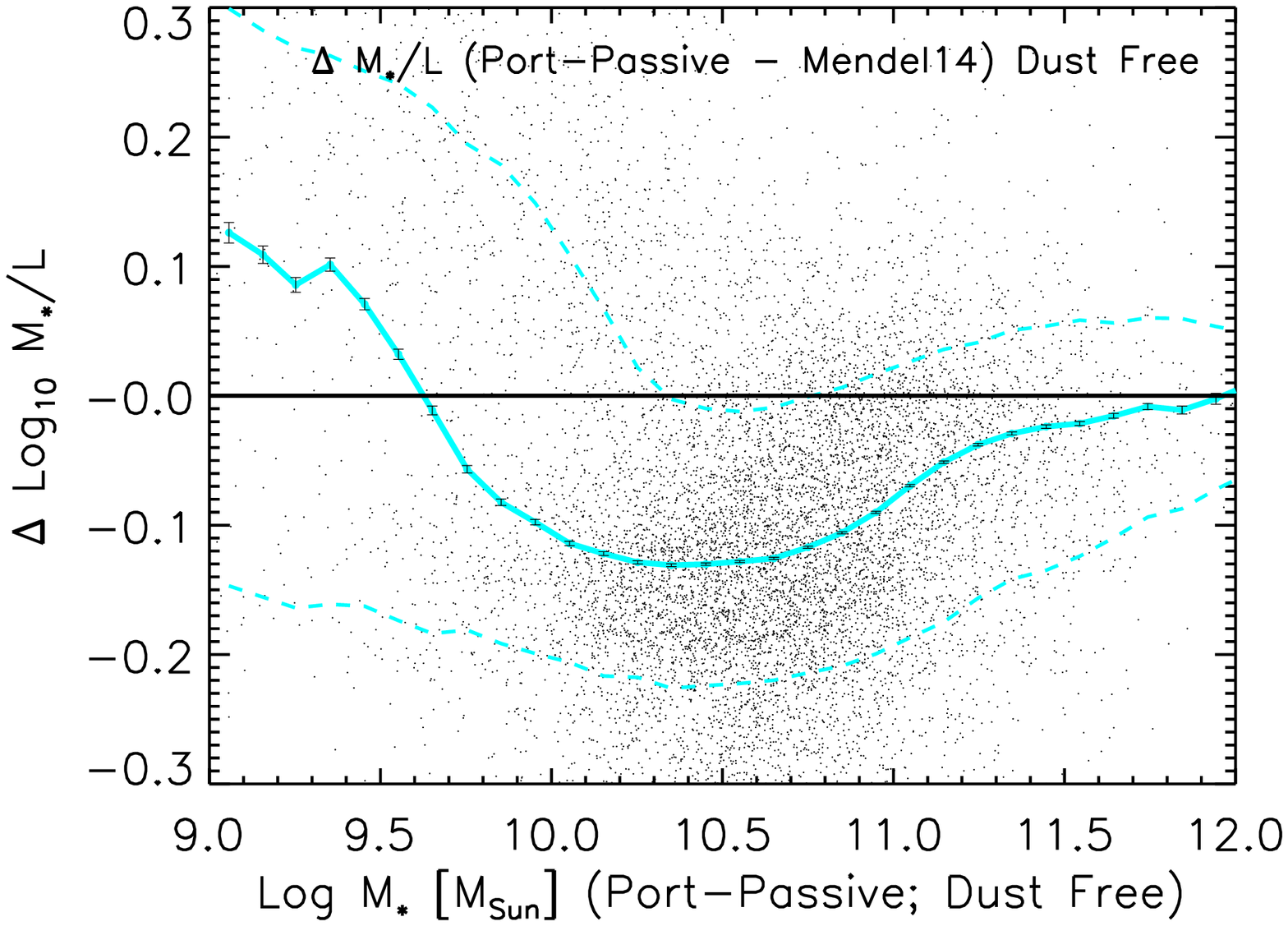}
 \caption{Comparison of the Portsmouth-Passive and Mendel et al. (2014) $M_*/L$ values for a Chabrier IMF.  (Both estimates assume galaxies are dust-free.) The disagreement at low masses should be ignored, because the Portsmouth-Passive models are only expected to be realistic at high masses $M_* >\! 10^{11}M_\odot$. Symbols and line styles same as in Figure~\ref{mendelDust}.}
 \label{mendelPort}
\end{figure}

The offset at large masses is part of the reason why Thanjavur et al. (2016) reported that their Mendel14-based $\phi(M_*)$ determination was smaller than that of B13.  Namely, their determination used dusty models.
Figure~\ref{mendelB13} shows that the offset between the Mendel14 dust-free $M_*/L$ values and those of B13 is of order 0.05~dex at high masses.  Combining this with Figure~\ref{mendelDust} shows that the difference in $M_*/L$ between B13 and the Mendel14 dusty models is of order 0.1~dex.  As Figure~\ref{thanja} in the main text shows, accounting for this offset (i.e., scaling the Mendel14 and B13 $M_*$ estimates to the same $M_*/L$ values) reduces the difference in $\phi(M_*)$ to less than 0.1~dex.  This indicates that the `fiducial' photometry of Thanjavur et al., which is based on Simard11, is very similar to the SerExp values of Meert15 which were used by B13, and which we use as the fiducial photometry in the main text.  In other words, the photometry agrees to better than 0.1~dex.  

Having shown the systematics associated with only changing the dust model, we now compare $M_*/L$ estimates when the treatment of the stellar population or the method of fitting the model to the data is varied.  Because we already know dusty and dust-free models differ, we do not compare the dusty models of one group with dust-free models of another.

We begin with a comparison of the MPA-JHU and Mendel14 models.  Figure~6 in Mendel et al. (2014) suggests that their $M_*/L$ values are in good agreement with the MPA-JHU values.  However, they did not account for the difference between their Chabrier and the MPA-JHU Kroupa IMFs.  Our Figure~\ref{mendelJHU} accounts for this difference (as one should), and this shows that the Mendel et al. values are actually about 0.05~dex larger than the MPA-JHU values. Some of the difference could be related to the inclusion of bursty models (MPA-JHU) or not (Mendel14). Above $10^{11}M_\odot$, the scatter around the median is $\sim 0.05$~dex.

Figure~\ref{mendelWisc} shows a similar comparison between the Wisconsin and Mendel14 values.  Comparison with the previous figure shows that the Wisconsin $M_*/L$ values are about 0.1~dex larger than the MPA-JHU values.  This is larger than the 0.05~dex scatter between the two models.  Finally, the top panel of Figure~\ref{mendelGranada} shows a similar comparison between the GranadaWide and Mendel14 values.  In this case, the scatter is substantially larger than in the previous two figures.  The reason why becomes evident in the bottom panel, which compares the dust-free versions of the two models.  The two clouds of points are approximately equally populated in the dust free models, but not so when dust is included.  It is the fact that there appear to be two populations which makes the scatter around the median so much larger.  The appearance of two populations is surprising, since both GranadaWide and Mendel14 are based on the same FSPS library, so the differences can only arise from details in how the library was used, and how the fitting was done.

Finally, for completeness, we compare the Portsmouth-Passive and Mendel14 values.  The disagreement at low masses should be ignored, because the Portsmouth-Passive models are only expected to be realistic at high masses.  Above $10^{11.3}M_\odot$ they are within 0.05~dex of the Mendel15 et al. values.

\end{document}